\documentclass[aps,prl,twocolumn,superscriptaddress,reprint]{revtex4-1}

\usepackage[ansinew]{inputenc}
\usepackage[T1]{fontenc}
\usepackage{ae,aecompl}
\usepackage[english]{babel}
\usepackage{subfigure,bm,dcolumn,graphicx,hyperref}
\usepackage{amsmath}
\usepackage{epstopdf}
\usepackage{color}
\usepackage[normalem]{ulem}

\begin{document}

\author{F. Arnold}
\altaffiliation{now at: Max Planck Institute for Chemical Physics of Solids, N\"othnitzer Str. 40, 01187 Dresden, Germany}
\author{A. Isidori}
\altaffiliation{now at: Scuola Internazionale Superiore di Studi Avanzati (SISSA), Via Bonomea 265, 34136 Trieste, Italy}
\affiliation{Royal Holloway, University of London, TW20 0EX Egham, United Kingdom}
\author{E. Kampert}
\affiliation{Hochfeld-Magnetlabor Dresden (HLD), Helmholtz-Zentrum Dresden-Rossendorf, D-01328 Dresden, Germany}
\author{B. Yager}
\author{M. Eschrig}
\author{J. Saunders}
\affiliation{Royal Holloway, University of London, TW20 0EX Egham, United Kingdom}

\title{Charge density waves in graphite: towards the magnetic ultra-quantum limit}

\date{\today}

\begin{abstract}
Graphite is a model system for the study of three-dimensional electrons and holes in the magnetic quantum limit, in which the charges are confined to the lowest Landau levels. We report magneto-transport measurements in pulsed magnetic fields up to 60~T, which resolve the collapse of two charge density wave states in two, electron and hole, Landau levels at 52.3 and 54.2~T respectively. We report evidence for a commensurate charge density wave at 47.1~T in the electron Landau level, and discuss the likely nature of the density wave instabilities over the full field range. The theoretical modeling of our results predicts that the ultra-quantum limit is entered above 73.5~T. This state is an insulator, and we discuss its correspondence to the ``metallic'' state reported earlier. 
We propose that this (interaction-induced) insulating phase supports surface states that carry no charge or spin within the planes, but does however support charge transport out of plane. 
\end{abstract}

\maketitle

\noindent 

Semimetals like graphite and bismuth are the subject of renewed interest due to their close relation to the topological Dirac and Weyl semimetals \cite{Liu14,Neupane14,Weng15}. Their low charge carrier density and effective mass reduce the magnetic field necessary to drive these systems into their magnetic quantum limit, in which the electronic structure is described by only a few Landau levels (LLs). In such materials this limit can be realized in state-of-the-art high-magnetic field laboratories, whereas it is inaccessible in conventional metals. The magnetic field induces a crossover from 3D to 1D physics, and the perfect nesting characteristic of 1D electron systems leads to the possibility of field-induced spin- and charge-density wave instabilities (SDWs and CDWs, respectively) with ordering wave-vector along the field direction \cite{Overhauser60,Overhauser68,Celli65,Halperin87}. In the ultra-quantum regime these systems are potentially subject to a rich variety of magnetic field driven topological quantum states such as fractional quantum Hall effect and the recently discussed axial and chiral anomaly \cite{Halperin87,Behnia07,Banerjee08,Yang10,Alicea14,Nielsen83,Kuechler14}. The clarification of the nature of the density wave instabilities in graphite, and the nature of the quantum state at high magnetic fields, is the focus of the work reported here.

\begin{figure}[b]
        \centering
                \includegraphics[width=1.0\columnwidth]{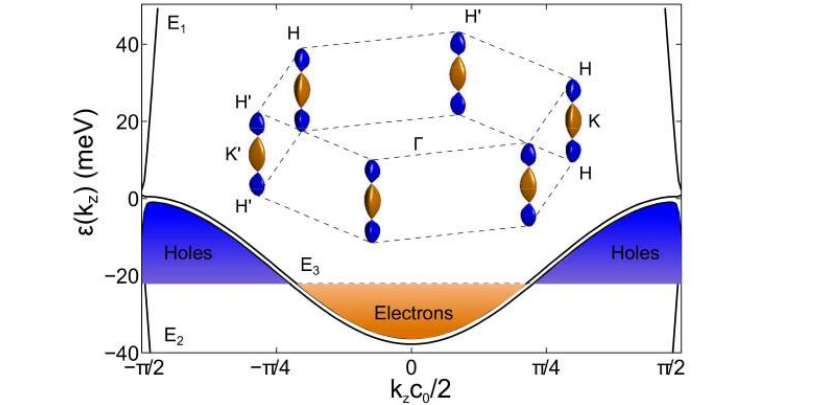}
        \caption{Zero-field $k_z$-dependence of graphite's band-structure along the H-K-H and H'-K'-H' edges of the Brillouin zone.}
        \label{fig:FigureA}
\end{figure}

Graphite is a 3D semimetal consisting of an infinite stack of graphene sheets \cite{Novoselov05,CastroNeto09}. The zero-field band structure is well described by the tight-binding Slonczewski-Weiss-McClure model \cite{Wallace46,McClure57,Slonczewski58,McClure60}. The Fermi surface consists of strongly anisotropic, trigonally warped ellipsoidal electron and hole pockets located along the H-K-H and H'-K'-H' edges of the hexagonal Brillouin zone (see Fig.~\ref{fig:FigureA}). 
In magnetic fields applied perpendicular to the graphene layers, 
the lowest-energy band ($E_3$, slightly hybridizing with two higher-energy bands $E_1$ and $E_2$ near the H points) gives rise to two 1D LLs with index $n=-1$ and $n=0$.
Each LL is spin-split by the Zeeman energy and has a residual twofold valley-degeneracy associated with the two inequivalent H-K-H and H'-K'-H' edges of the Brillouin zone. For magnetic fields above 8 T these are the only LLs that cross the Fermi energy. The system becomes quasi-1D, due to c-axis dispersion.

\begin{figure*}[tb]
        \centering
                \includegraphics[width=1.0\textwidth]{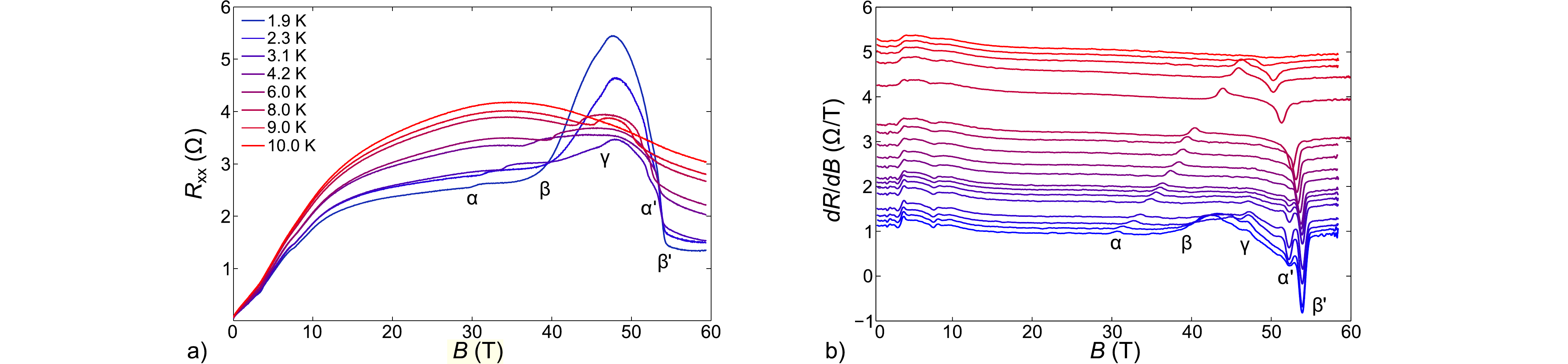}
        \caption{a) Magnetic field dependence of the in-plane resistance of graphite for temperatures below 10 K measured in a 25 ms 60 T pulsed field magnet. b) Magnetic field derivatives of the in-plane resistance: The data has been offset proportional to the temperature by 0.5 $\Omega/\mathrm{T K}$ for clarity. $\alpha$ and $\beta$ denote the onset transitions of the DWs in the ($0,\uparrow$) and ($-1,\downarrow$), whereas $\alpha'$ and $\beta'$ mark the field at which they vanish. At $\gamma$ the CDW associated with the ($0,\uparrow$) LL undergoes a lock-in transition.}
        \label{fig:FigureCG}
\end{figure*}

This quantum limit has been extensively investigated by magneto-transport measurements. Studies on Kish graphite showed a pronounced resistance anomaly above 22 T \cite{Tanuma81,Iye82}. Yoshioka and Fukuyama \cite{Yoshioka81} attributed the observed anomaly to the formation of a CDW in one of the lower LLs \cite{Nakao76}, and developed a mean-field theory describing the magnetic field dependence of its critical temperature. Improved LL calculations by Takada and Goto \cite{Takada98} showed that electron correlations play a crucial role in renormalizing the Slonczewski-Weiss-McClure LL structure. Subsequent experiments to higher fields \cite{Yaguchi98,Yaguchi09} identified a transition at 52~T corresponding to the collapse of a density wave phase as the ($0,\uparrow$) or ($-1,\downarrow$) LL emptied. So far attempts to directly measure the gap of this putative density wave spectroscopically \cite{Latyshev12,Arnold13} have been unsuccessful. 

Recently, Fauqu\'e et al. \cite{Fauque13} discovered the onset of a second density-wave anomaly above 53~T,  by out-of-plane transport measurements on Kish graphite up to 80 T. This state was found to collapse at 75~T into a state with metallic c-axis conductivity. These authors argued that only one of the ($0,\uparrow$) or ($-1,\downarrow$) LLs depopulates at 53~T, and furthermore that the ($0,\downarrow$) and ($-1,\uparrow$) levels remain populated above 75~T. On the other hand, the observation of a vanishing Hall coefficient at 53~T \cite{Akiba15} was interpreted in terms of the depopulation of both the ($0,\uparrow$) and ($-1,\downarrow$) LLs at 53~T, as predicted theoretically \cite{Takada98}. It was suggested that an excitonic phase forms in the remaining ($0,\downarrow$) and ($-1,\uparrow$) levels above 53~T . Elsewhere it has been argued that an excitonic insulator phase appears at 46~T \cite{Zhu15}. A resistance hysteresis was observed around 53 T \cite{Yaguchi13}. In addition, a small feature of unclear origin (we will clarify its origin in the present work) around 47 T was reported \cite{Yaguchi93}.
It is clear that the evolution of the LLs with magnetic field, and their associated density wave instabilities, through the quantum limit and into the ultra-quantum limit at highest fields, remain open questions.

\begin{figure}[b]
        \centering
        \includegraphics[width=1.0\columnwidth]{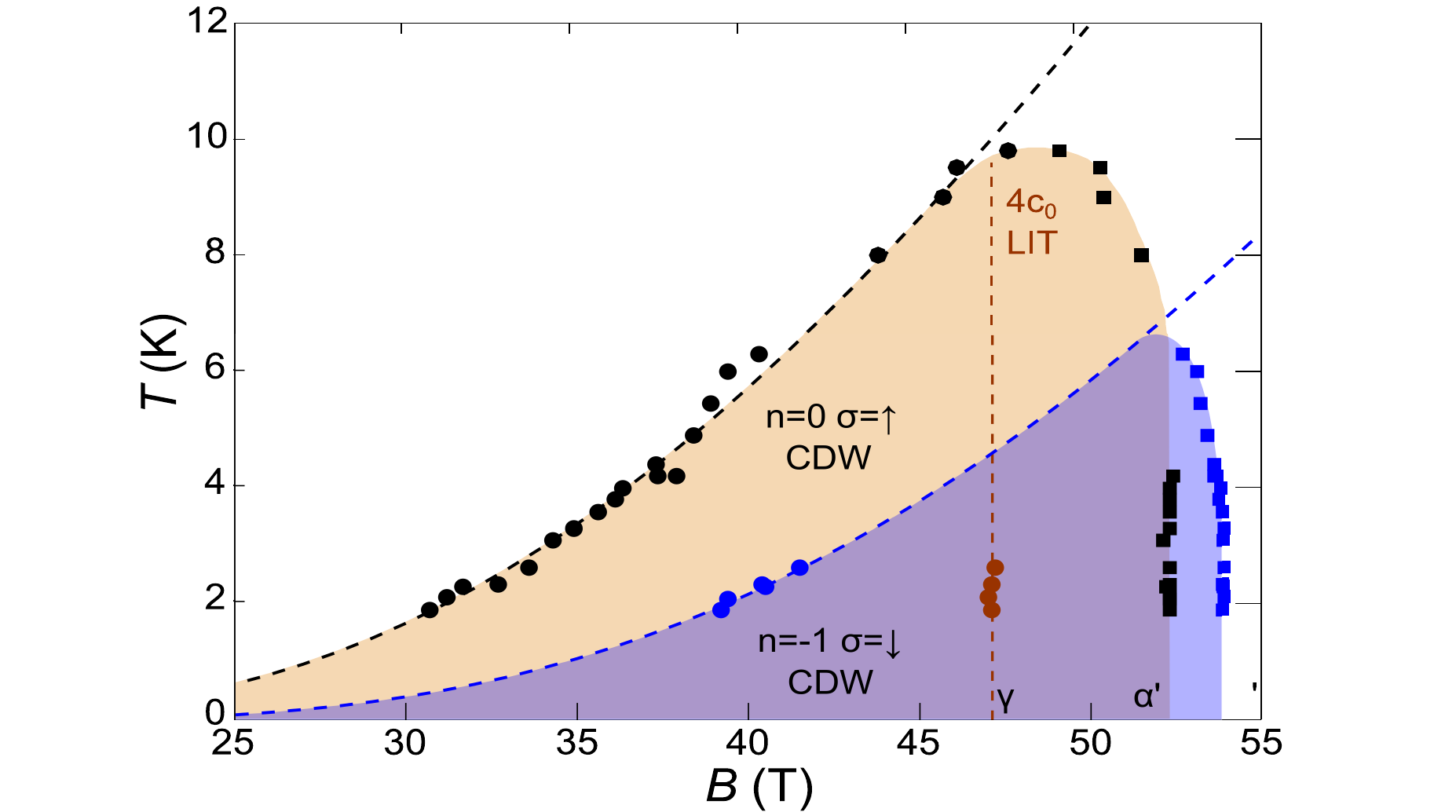}
        \caption{Phase diagram of the CDWs in the ($0,\uparrow$) and ($-1,\downarrow$) LLs. Circles and squares are phase transitions with positive and negative differential magnetoresistance, where black represents the $\alpha$, blue the $\beta$, and ocher the $\gamma$-transition. At low fields the critical temperature of $\alpha$ and $\beta$ increases exponentially following \cite{Yoshioka81} (black and blue dashed lines): $T_c(B) = T^* \exp\{-B^*/B\}$, with $T^*_\alpha=230$~K, $B^*_\alpha=148$ T, $T^*_\beta=300$ K, $B^*_\beta=195$ T.}
        \label{fig:Figure3}
\end{figure}
In this article we present new magneto-transport measurements on single crystal graphite, of significantly higher quality than hitherto, in conjunction with new theoretical calculations of the LL band structure, and density wave instabilities. We show that the field-induced density wave below 54.2 T, contrary to previous reports, is a superposition state of a number of incommensurate collinear CDWs with different onset and vanishing fields. One of these undergoes a lock-in transition at 47.1 T before two of the CDWs vanish due to the emptying of their corresponding LLs at 52.3 and 54.2 T respectively. 
Within our model we predict that the system stays gapped at low temperatures above 54.2 T, due to additional CDW's which empty ultimately at 73.5 T. This upper threshold field defines the ultra-quantum limit, above which our model predicts a LL structure such that the bulk is an insulator.
This corresponds to the new phase, with observed onset at 75~T, discovered in \cite{Fauque13}. Given the observation of metallic c-axis response in that phase \cite{Fauque13}, we propose that it supports a set of surface states which allow for charge transport perpendicular to the planes, although within the planes the charge- and spin currents cancel. We find that interactions are crucial for this state to be insulating; without taking into account interactions the system would be metallic at these magnetic fields.

\begin{figure*}[tb]
        \centering
                \includegraphics[width=1.0\textwidth]{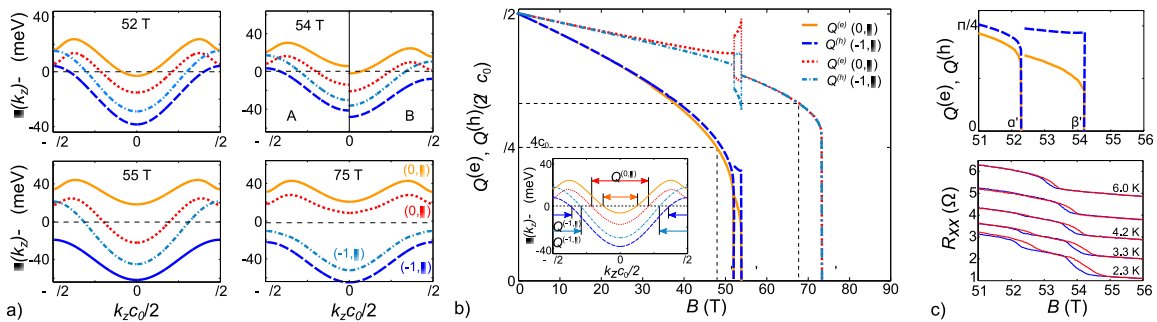}
        \caption{a) Renormalized LL structure in the ultra-quantum limit: Above 8 T only four LLs cross the Fermi energy. At 54~T valley degeneracy is lifted leading to two differentially doped sets of LLs, here A and B refer to the H-K-H and H'-K'-H' valleys. b) CDW nesting vectors $Q$ up to 90 T. Hole-like levels have nesting vectors given by $Q^{\rm (h)}=2 k_F^{\rm (h)} = 2 \pi/c_0 - 2 k_F$ (see inset). All $Q$'s decrease with increasing magnetic field due to the upward and downward shift of the electron-like ($0,\sigma$) and hole-like ($-1,\sigma$) LLs, before vanishing in a first-order transition at $B_{\alpha'}=52.3$ T, $B_{\beta'}= 54.2$ T, and $B_{\delta'}=73.5$ T. The nesting vector of the ($0,\uparrow$) LL becomes commensurate at $B_\gamma=48$ T. c) Close-up of the nesting vectors in the vicinity of the first-order $\alpha'$ and $\beta'$-transitions (top panel) and observed resistance hysteresis (bottom panel): The red curves show the magnetic field up-sweeps whereas blue curves correspond to the down-sweeps.}
        \label{fig:Figure4}
\end{figure*}

Our electrical transport measurements were performed on a single crystal of Tanzanian natural graphite \cite{NaturallyGraphite}, the quality of which significantly exceeds the commonly used Kish graphite, highly-oriented-pyrolytic graphite, or other natural graphites, as evidenced by de Haas-van Alphen measurements \cite{SOM}. This sample quality is crucial in revealing the new features we observe; the self-doping of graphite by structural defects is minimised, reducing electron-hole charge imbalance. Our studies of the in-plane and out-of-plane magnetotransport were made in pulsed magnetic fields up to 60 T \cite{Zherlitsyn10,SOM}.

Figure \ref{fig:FigureCG} shows the in-plane resistance of the graphite crystal for fields up to 60 T and temperatures below 10 K. The in-plane resistance first increases steeply, superimposed by Shubnikov-de Haas oscillations, and saturates above 15 T due to the presence of a closed Fermi surface \cite{Brandt75,Sugihara78,Iye82,Halperin87}.  At lowest temperatures a step is observed in the in-plane resistance at around 30 T, followed by a steep increase of the resistance, reaching its maximum value at 48 T before it drops down to below its initial value at around 53 T. This behavior has already been reported for Kish and highly-oriented pyrolytic graphite and has been attributed to the formation of a DW state \cite{Iye82,Yaguchi93,Uji98,Yaguchi09,Fauque13}. Similar features can also be identified in the out-of-plane transport \cite{SOM,Fauque13}.

In order to highlight the signatures of the transitions in our sample, the magnetic field derivatives of the in-plane magnetoresistance were calculated (see Fig. \ref{fig:FigureCG}). Here the transitions appear as maxima and minima. We denote the observed maxima as $\alpha$, $\beta$, and $\gamma$, in ascending order of their transition fields, and the corresponding minima as $\alpha'$ and $\beta'$. The resolution of the two features $\alpha'$ and $\beta'$ is the first key new observation, and 
we also observe it in the out-of-plane transport \cite{SOM}. The $\beta'$ transition develops below 6 K alongside the $\alpha'$ transition. In addition we identify a new weaker feature, labelled $\gamma$, at $47.1\pm0.1$~T, which is resolved below 3 K and represents the onset of another resistance increase. 
The temperature dependence of these transitions is shown in Fig. \ref{fig:Figure3} \cite{SOM}.

We now discuss the theoretical interpretation of these features, based on our new numerical calculations of the LL structure including the Hartree-Fock self-energy, following the framework of \cite{Takada98}. In the quantum limit, the system is understood in terms of the one-dimensional dispersion of the lowest, spin-split, electron and hole LLs, and their density wave instabilities.  We investigate closely the magnetic field dependence of the putative nesting vectors in all four LLs that are close to the Fermi level (see Fig. \ref{fig:Figure3}). We calculate the renormalized LLs by including the electron self-energy in a fully self-consistent manner \cite{SOM}.
The effect of short-range correlations, neglected by the conventional random-phase approximation, is included via the Hubbard local-field correction \cite{Hubbard57,Singwi68} to the effective electron-electron interaction. The inclusion of this correction is crucial in order to get a quantitative agreement between the theoretical LL structure and the experimental data. We use the experimental value of $B_{\alpha'}$ to fix the tuning parameter that characterizes our theory \cite{SOM}, namely the static relative permittivity.

We find that the sharpest new features at $B_{\alpha'}=52.3\pm0.1$ T and $B_{\beta'} = 54.2\pm0.1$ T correspond to two distinct first-order transitions (see Figs.~\ref{fig:Figure3} and \ref{fig:Figure4}) associated with the abrupt depopulation of both the $(0,\uparrow)$ (electron-like) and $(-1,\downarrow)$ (hole-like) LLs.
We argue below that the splitting of the transitions arises from the breaking of valley degeneracy. The first-order character of the $\alpha'$ and $\beta'$ transitions, predicted by our theory, is experimentally confirmed by the clear evidence of hysteresis in both, as shown in Fig.~\ref{fig:Figure4}c.

Furthermore, our calculations show that at $B_{\gamma}=48$~T and $B_{\epsilon}=68$~T the nesting vectors of the intra band CDWs, i.e. $Q^{(0,\uparrow)}$ and $Q^{(-1,\downarrow)}$ as well as $Q^{(0,\downarrow)}$ and $Q^{(-1,\uparrow)}$ (Fig. \ref{fig:Figure4}b) take the values of $2\pi/4c_0$ and $2\pi/3c_0$ respectively, corresponding to commensurate wavelengths of $4c_0$ and $3c_0$ (the lattice parameter $c_0$ is twice the distance between adjacent graphene layers). This suggests the $\gamma$-feature, observed experimentally around 47.1 T and temperatures below 3 K (see Fig.~\ref{fig:FigureCG}) is a lock-in transition of the corresponding CDW from an incommensurate state to a ``phase-locked'' commensurate order \cite{Lee74,Moncton75,McMillan76,Bak76,Bruce78b,Bruce78c,Cowley80,Sugai06,SOM}.  Lock-in transitions are characteristic fingerprints of CDW systems \cite{Monceau89}, where translational invariance of the charge modulation is lost with respect to the lattice \cite{SOM}.

We now elaborate on the splitting of the $\alpha'$ and $\beta'$ transitions. 
In the simplest calculation these transitions occur at the same field value, in an undoped sample \cite{doping}. To account for the splitting, while preserving overall charge balance, we need to either lift the valley degeneracy or invoke an inhomogeneous state with spatial phase separation. In the first scenario, a spontaneous breaking of the valley degeneracy leads to a splitting between the LLs originating from the Fermi surface pockets located along the two inequivalent H-K-H and H'-K'-H' edges of the Brillouin zone (see Fig.~\ref{fig:FigureA}) similar to what has been observed in bismuth \cite{Kuechler14}. The system is then characterized by two inequivalent sets of LLs, one for each valley, allowing each set of LLs to develop a finite, equal and opposite, charge imbalance in the narrow region between $B_{\alpha'}$ and $B_{\beta'}$. In this magnetic-field region both valleys are characterized by three populated and one empty LLs; the empty level in the hole-doped valley is the electron-like $(0,\uparrow)$ LL, whereas in the electron-doped valley the empty level is the hole-like $(-1,\downarrow)$ (see Fig. \ref{fig:Figure4}). Since the two valleys are now characterized by inequivalent nesting vectors, the collinear CDW order observed below $\alpha'$ turns into a valley density-wave state, with an additional commensurate in-plane modulation originating from the reciprocal lattice vectors connecting the two valleys. We find that such a state is promoted by moving the system towards another lock-in state \cite{SOM}.

To account quantitatively for the observed splitting between $\alpha'$ and $\beta'$ we predict a charge imbalance, parameterising the broken valley degeneracy, of $(p-n)/(p+n) = \pm 0.128$ in the two valleys, where $p$ ($n$) denotes the hole (electron) carrier density. The opposite imbalance in the two valleys guarantees the overall charge balance of the system.

We now turn to the ultra-quantum limit. We first note that, for magnetic fields above $\beta'$, the populated levels are $(0,\downarrow)$ (electron-like) and $(-1,\uparrow)$ (hole-like). These LLs are still characterized by finite nesting vectors. Note that the nesting vectors arising from these bands are identical due to the charge balance between electron and hole carriers. 

Our calculations show that both levels become depopulated at 73.5~T (see Fig.~\ref{fig:Figure4}). This state was first theoretically identified by \cite{Takada98} in a computation at 200~T.
Our predicted numerical value of $B_{\delta'}$ is in remarkable agreement with the measured critical field of 75 T for the reentrant transition of the upper density-wave phase observed in \cite{Fauque13}. Furthermore we predict that the nesting vectors of a putative CDW in the $(0,\downarrow)$, $(-1,\uparrow)$ levels  gives rise to an additional lock-in transition with a wavelength of $3c_0$ at $68~\mathrm{T}$. Observation of this transition would conclusively identify the DW instabilities in this field range as CDWs.

Our finding that the $(0,\downarrow)$ and $(-1,\uparrow)$ states empty above 75T refutes the proposed energy level scheme to account for the observed metallic conductivity in this field regime, in Ref. \cite{Fauque13}, where it is claimed that the $(0,\downarrow)$ and $(-1,\uparrow)$ states remain occupied. In contrast, above 75T we find that the bulk should be fully gapped.

We speculate that low temperature transport may be expected to occur via a 2D chiral sheath of extended surface states \cite{Fauque13,MacDonald90,Chalker95,Balents96,Cho97,Druist98,Young14}. These surface states are protected against weak impurity scattering. This leads to a ballistic in-plane transport along the edges of the sample, in the direction perpendicular to the magnetic field. Taking into account valley degeneracy, all four LLs contribute, leading to a cancellation of the charge and spin currents. Nevertheless metallic conductivity is observed in the out-of-plane direction \cite{Fauque13}, and we propose that this is accounted for by inter-layer hopping occurring at the sample edges \cite{Chalker95,Balents96,Cho97}. At magnetic fields between  55 and 75 T the Fermi energy crosses the ($0,\downarrow$) and ($-1,\uparrow$) LLs. The wavefunctions of surface states have finite overlap with bulk states and are scattered by bulk impurities, which renders transport via these states diffusive, yielding a thermally activated out-of-plane conductance at temperatures below the DW transition, as observed \cite{Fauque13}. Studies of in-plane transport in this field regime should be of great interest.

From our assignment of the Landau level structure, it appears that in the ultra-quantum limit graphite above 75 T is a fully gapped state in the bulk, an insulator, due to the strong electron correlations on the LL band structure. This new state potentially harbors an abundance of new physics and is experimentally accessible in state-of-the-art pulsed magnetic field facilities. There is current interest in systems where emergent band-structure leads to a topologically non-trivial insulating state, topological Kondo insulators being one example \cite{Dzero,Dzero2}. The observation of metallic out-of-plane transport \cite{Fauque13} opens a new play-ground for studying surface states in interaction-induced insulators, with the possibility of a topological phase transition at fields beyond $75\,\mathrm{T}$.

In conclusion the direct observation, in high quality samples, of the depopulation of two LLs at $52.3$ and $54.2~\mathrm{T}$, and the commensuration of the DW at $48~\mathrm{T}$ imposes severe constraints on the nature of the field-induced DW instabilities in graphite. We can theoretically account for these features, and correctly predict the field quenching of the DWs above 54 T, observed recently \cite{Fauque13}. We believe that the most likely scenario at $B<B_{\alpha'}$ and $T=0$ is CDWs in each of the four LLs. A full discussion of all possible nesting instabilities and DW scenarios and the logic behind this proposal is presented in \cite{SOM}, where the signatures from in-plane and out-of-plane transport are also discussed. A test of this hypothesis would be the observation of the lockin transition we predict at $68~\mathrm{T}$.

\begin{acknowledgments}
The authors wish to thank T. F\"orster, J. Wosnitza, T. Herrmansd\"orfer and S. Zherlitsyn for experimental support at the HLD.  We acknowledge the support of the HLD-HZDR, member of the European Magnetic Field Laboratory (EMFL) as well as the Hubbard Theory Consortium, the Max Planck Society (MPRG Physics of Unconventional Metals and Superconductors, E. Hassinger) and the Engineering and Physical Science Research Council (EPSRC Grant Nos. EP/H048375/1 and EP/J010618/1).
\end{acknowledgments}

\clearpage

\section*{Supplemental Material for ``Charge density waves in graphite: towards the magnetic ultra-quantum limit''}

\subsection{1. Sample Quality and Preparation}

\begin{figure}[b]
        \centering
                \includegraphics[width=0.90\columnwidth]{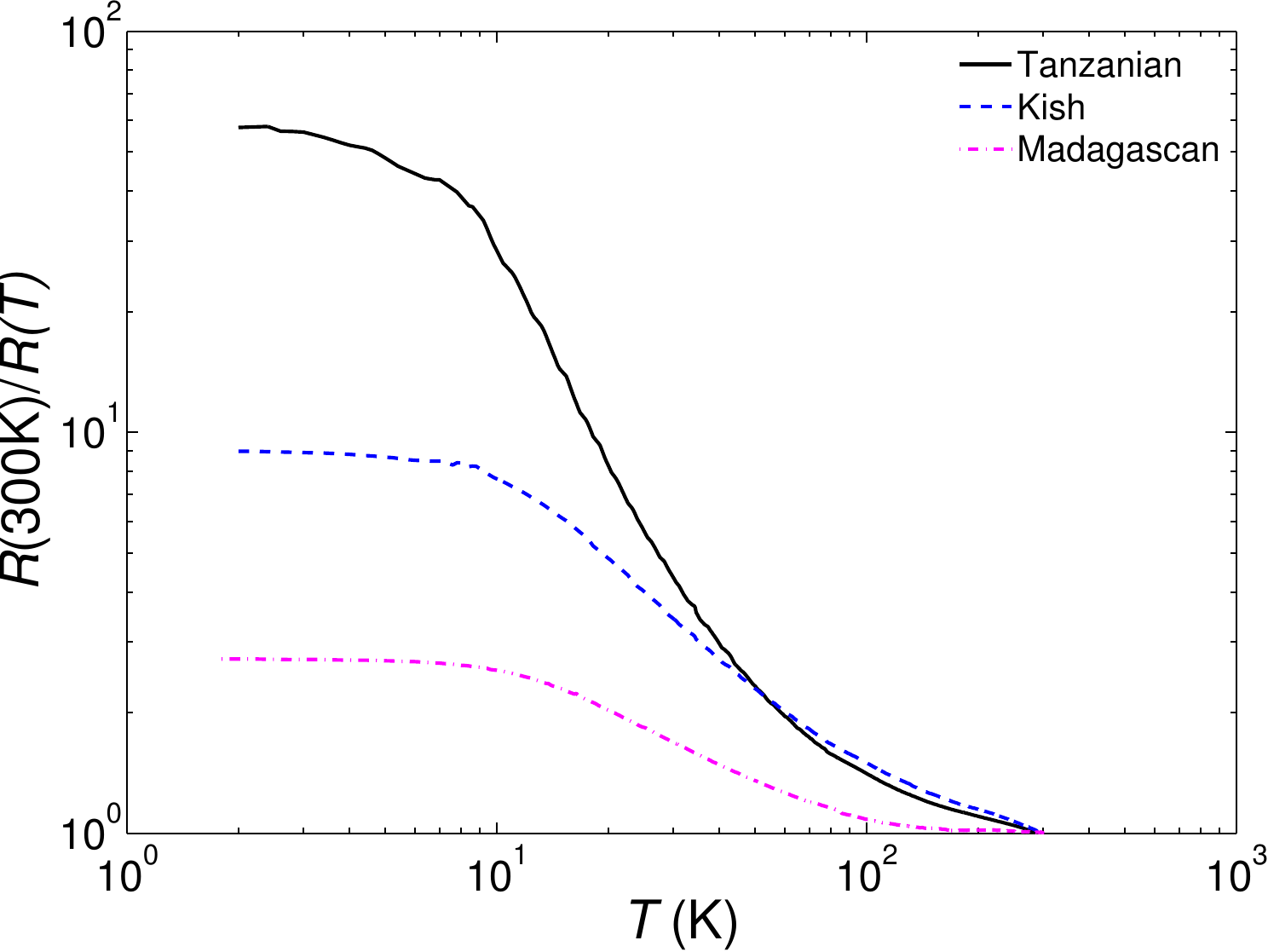}
        \caption{Temperature dependence of the in-plane conductivity of various single crystalline graphites at zero magnetic field, normalized by their room temperature value. Our sample (solid line) has a much higher residual resistivity ratio than other single-crystal graphites.}
        \label{fig:Figure1SOM}
\end{figure}

\noindent The samples studied in this article are natural Tanzanian graphite single crystals \cite{NaturallyGraphite}. Tanzanian graphite offers superior sample quality compared to other forms of natural and artificial graphite. Our sample shows a residual resistivity ratio (RRR) of 60, which exceeds the typical RRRs of Kish(10), Madagascan(2.5), and highly-oriented-pyrolitic graphite(<10) (see Fig~\ref{fig:Figure1SOM}). 

Furthermore we prove the better sample quality by a comparative Dingle analysis. Here we find a lower Dingle temperature and consequentially higher quantum oscillation amplitude. For this the magnetic field dependences of the de Haas-van Alphen amplitude were studied at $2\,\mathrm{K}$ in magnetic fields up to $7\,\mathrm{T}$ using a Quantum Design SQUID VSM. Samples were oriented such that the magnetic field was applied parallel to the crystallographic c-axis.
\begin{figure}[tb]
	\centering
		\includegraphics[width=1.0\columnwidth]{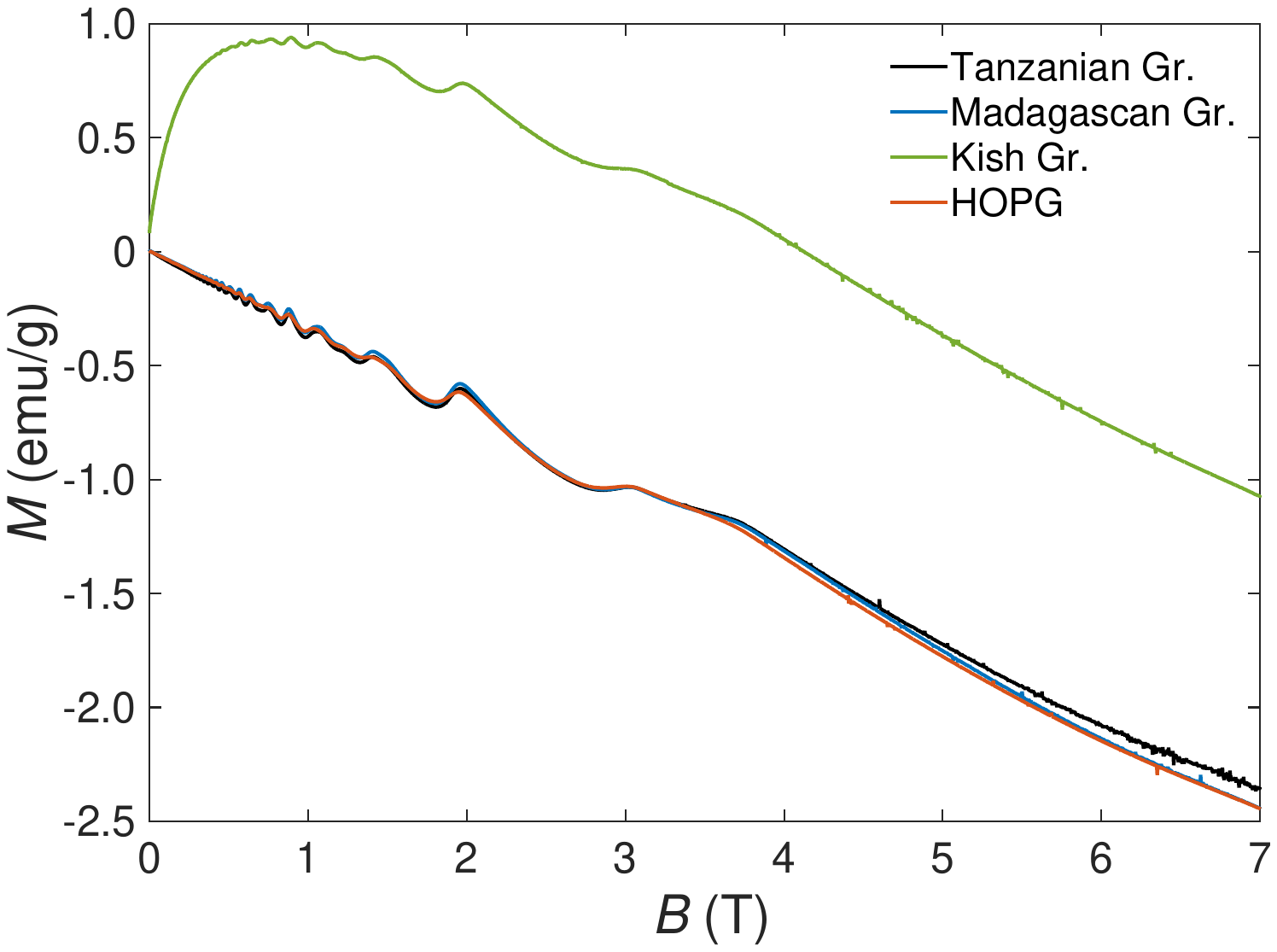}
	\caption{Specific magnetization of different types of graphite  measured at $2\,\mathrm{K}$.}
	\label{fig:Magnetisations}
\end{figure}

\begin{figure}[tb]
	\centering
		\includegraphics[width=1.0\columnwidth]{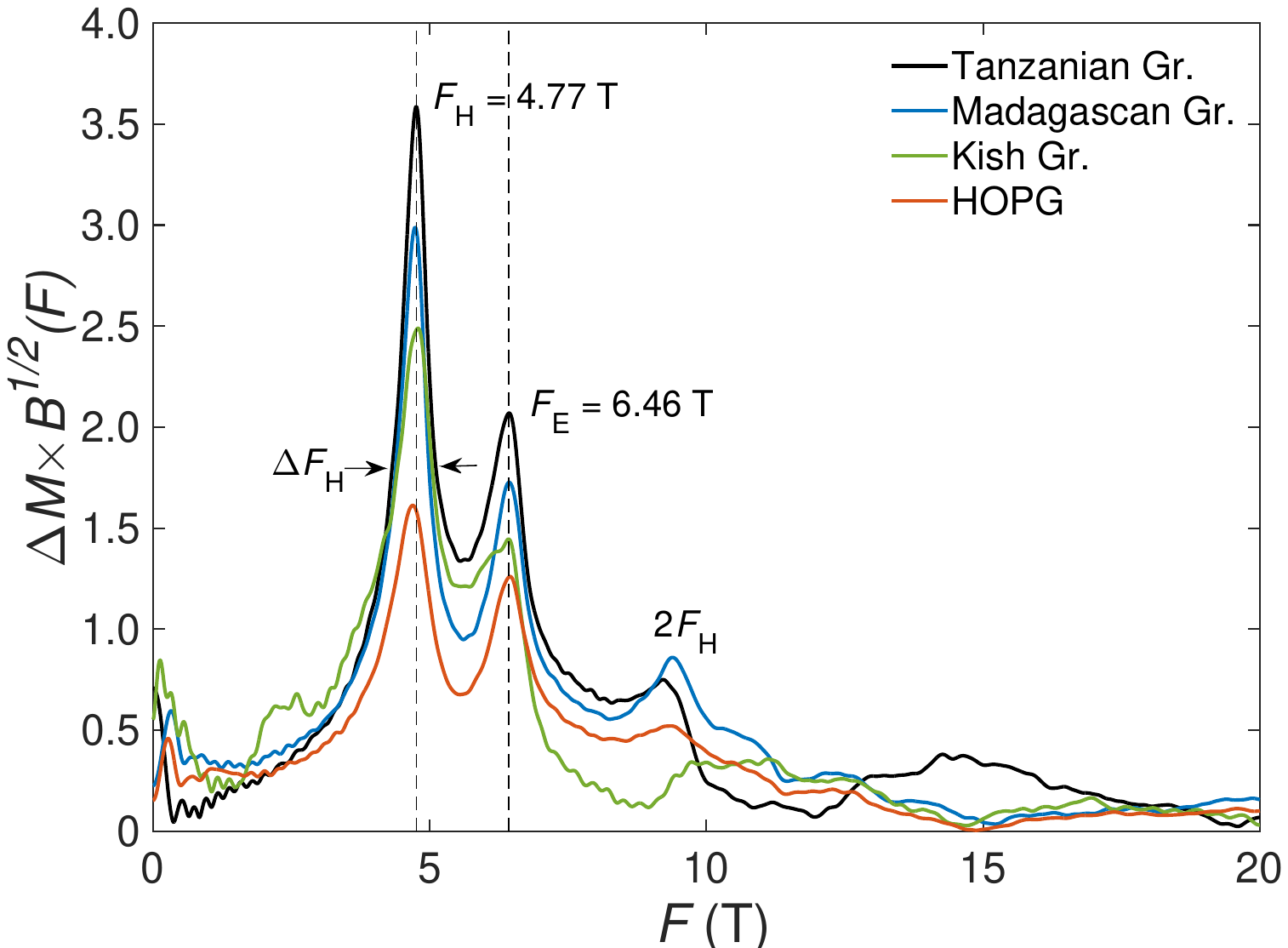}
	\caption{Fourier transforms of the background subtracted and field corrected de Haas- van Alphen signal of different types of graphite. The Fourier transforms were taken in the magnetic field interval from 0.2 to $2.5\,\mathrm{T}$ based on the specific magnetization shown in Fig. \ref{fig:Magnetisations}.}
	\label{fig:FourierTransforms}
\end{figure}

Figure \ref{fig:Magnetisations} shows the specific magnetization of Tanzanian, Madagascan, Kish and highly-oriented-pyrolitic graphite. Due to the presence of two similar quantum oscillation frequencies originating from the electron and hole pockets, a strong beating of the quantum oscillations is observed and induces a large error when fitting the envelope function. Thus we determine the Dingle temperatures from the Fourier transforms of the magnetic field corrected de Haas-van Alphen signal $\Delta M\sqrt{B}(1/B)$ (see Fig. \ref{fig:FourierTransforms}). Here the exponentially decaying Dingle term transforms to a Lorentzian line shape, where the full-width half maximum $\Delta F=2\pi k_\mathrm{B} T_\mathrm{D} m^*/\mu_\mathrm{B}$ \cite{Arnold16NatCom}.

The Dingle temperature can be further used to calculate the de Haas-van Alphen charge carrier mobility $\mu = e\tau/m^*$ where the scattering time $\tau=\hbar/2\pi T_\mathrm{D}$.

Based on the line width of the hole pocket signal $\Delta F_\mathrm{H}$ in Fig. \ref{fig:FourierTransforms}, we calculate the hole Dingle temperatures and mobilities $\mu_\mathrm{H}$ shown in Tab. \ref{tab:DingleAnalysis}.

\begin{table}[tb]
\caption{Dingle temperatures and charge carrier mobilities of graphites}
\begin{tabular}{l | c  c  c}
Sample & $\Delta F_\mathrm{H}$ (T)& $T_\mathrm{D} (K)$ & $\mu_\mathrm{H} (cm^2/Vs)$\\
\hline
Tanzanian & 0.79& 4.33 &$1.27\times10^4$\\
Madagascan & 0.81& 4.41 &$1.24\times10^4$\\
Kish & 1.36& 7.45 &$7.4\times10^3$\\
HOPG & 1.23& 6.77 &$8.1\times10^3$
\end{tabular}
\label{tab:DingleAnalysis}
\end{table}

Due to overlap with the dominant hole pocket line, we were not able to extract Dingle temperatures from the subdominant electron pocket signal. 

In addition to the highest RRR, we do also find the lowest Dingle temperature and highest quantum oscillation amplitude in Tanzanian graphite, thus proving its outstanding quality.

Graphite crystals were cut in the dimensions $3\,\times0.5\,\times0.080$ mm and $0.5\,\times0.5\,\times0.150$ mm for in- and out-of-plane transport samples respectively. For in-plane measurements, electrical contacts were made in standard six-point geometry to the top-surface of the samples by gluing $25$ $\mu\mathrm{m}$ gold wires to the sample using silver loaded epoxy (EpoTek H20E and Dupont 6838). These contacts were subsequently cured for 20 minutes at $120^o\mathrm{C}$. Out-of plane resistances were measured in four-point geometry, where a current and voltage contact were glued to the top and bottom surface of the sample.

\subsection{2. Pulsed Field Electrical Transport Measurements}

\noindent High magnetic field electrical transport measurements were performed at the Dresden High Magnetic Field Laboratory in a KS21 type pulsed field magnet. In this setup magnetic fields up to 60 T are achieved by discharging a capacitor array through a liquid nitrogen cooled normal resistive coil \cite{Zherlitsyn10}. The KS21 magnets produce magnetic field pulses with an overall pulse length of 25 ms. The magnetic field rises in 6 ms to its maximum value of 60 T and falls subsequently back to zero within 14 ms. The calibration error of the field pick up coil at the highest fields is about 200 mT. 

The high field sweep rates of up to 15 kT/s at the beginning of the pulse can cause eddy-current heating in highly conductive samples leading to a temperature discrepancy between the magnetic field up- and down-sweep. Thus the temperature resolution of the experiments is limited to about 0.2 K.

Resistances were measured with a constant excitation current of 2 to 5 mA and frequency of 33 to 72 kHz. Voltage and current reference signals were amplified with Stanford Research SR560 preamplifiers and recorded with a 1 MS/s, 16 bit Yokogawa DL750 oscilloscope. The sample resistances were calculated by applying a digital lock-in procedure to the recorded data. To account for Hall resistances the resulting resistances were averaged over magnetic field pulses with positive and negative polarity.

\subsection{3. Out-of-plane Resistance}

\begin{figure}[tb]
    \centering
    \includegraphics[width=1.00\columnwidth]{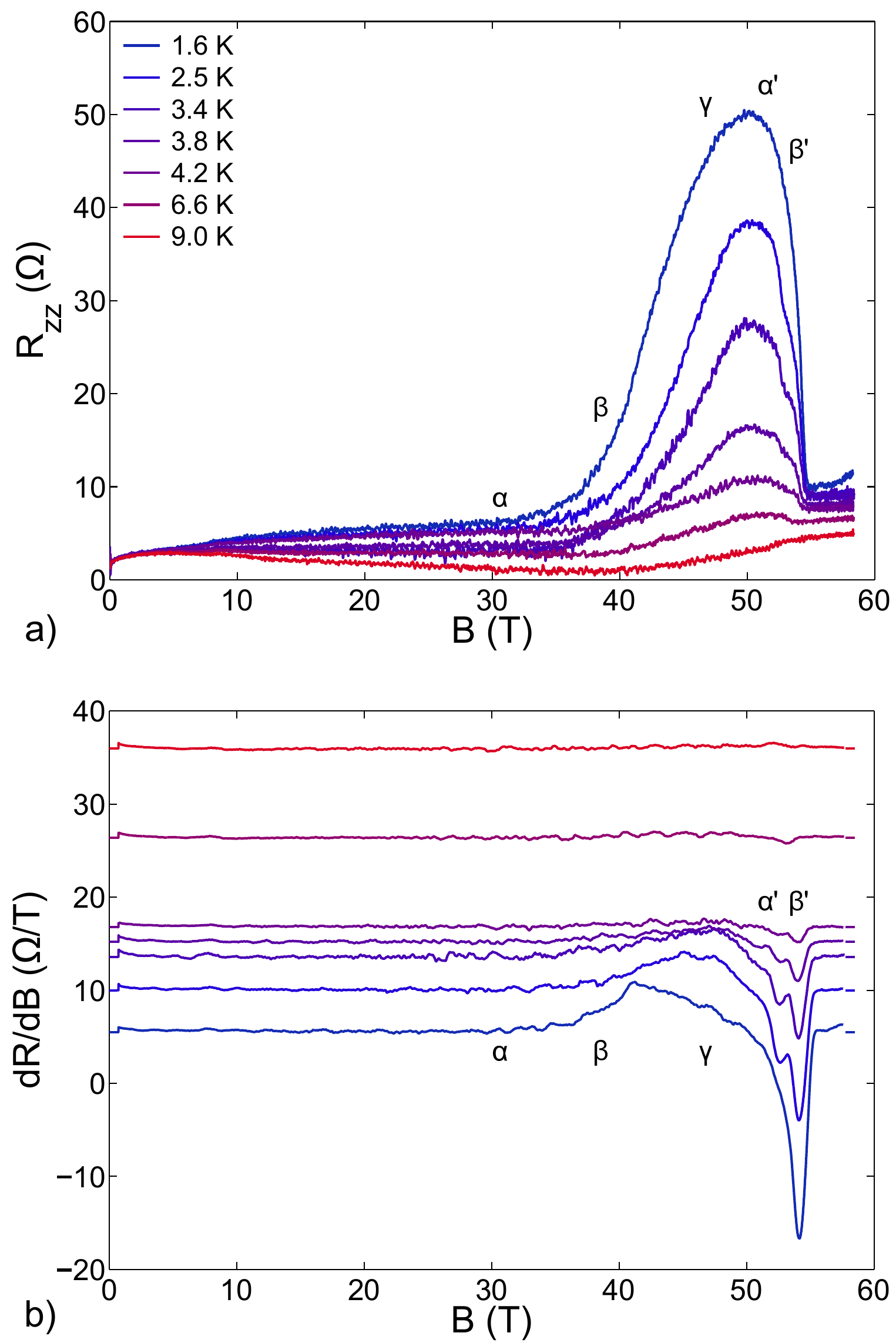}
\caption{Magnetic field dependence of the out-of-plane resistance (a) below 10 K, and corresponding differential magnetoresistance (b). Differential magnetoresistance curves are offset proportional to the temperature they were measured at. The resistance along the magnetic field direction is only weakly magnetic-field dependent up to the onset transition $\beta$. As in the in-plane transport, the out-of-plane resistance shows a pronounced anomaly above 30 T with the same $\beta$, $\alpha'$ and $\beta'$ signatures. The features associated with $\alpha$ and $\gamma$, however, are absent or strongly suppressed.}
        \label{fig:Figure2SOM}
\end{figure}

\noindent As discussed in the main text, measurements of the out-of-plane resistance confirm most of the phase-transitions observed in the in-plane resistance. As shown in Fig.~\ref{fig:Figure2SOM}, below the onset field of the first density wave (DW) instability the out-of-plane magnetoresistance is only weakly field- and temperature-dependent. However above 30 T a strong resistance anomaly develops similar to the one observed in the in-plane resistance. At the $\beta$-transition the out-of-plane resistance increases almost exponentially with magnetic field and decreases sharply at the $\alpha'$ and $\beta'$-transitions. The characteristic field dependence and critical fields are in good agreement with the data obtained from the in-plane measurements, as can also be seen in the differentiated magnetoresistance (bottom panel of Fig.~\ref{fig:Figure2SOM}). 

Compared to data reported earlier \cite{Fauque13}, we note that our resistance increase when entering the DW dome is much suppressed. The poor aspect ratio of our c-axis transport sample ($l/w\approx0.3$) combined with the particular 4-point contact geometry and large inverted resistance anisotropy at high magnetic fields leads to a decoupling of the current path from the voltage contacts. This effect known as "`current jetting"' is present in high mobility metals where the transverse resistivity, perpendicular to the magnetic field, increases strongly compared to the longitudinal resistivity \cite{Pippard,dosReis16}. In such a case the homogenously distributed current at zero field concentrates along a narrow path between the current electrodes at high fields. This current redistribution  leads to the observed negative magnetoresistance up to 40 T at high temperatures and reduced DW resistance at low temperatures.

In comparison to the in-plane transport, the signatures of the $\alpha$ and $\gamma$ transitions are absent or strongly suppressed in the out-of-plane data. The absence of the $\alpha$-signature in the out-of-plane transport has already been discussed by Iye and Dresselhaus \cite{Iye85}. However a more evolved explanation based on field induced magnetoresistance anisotropy and subdominant DW gaps can be found in Sec. 9 of this Supplemental Material.

\subsection{4. Lock-in Transition}

\noindent As thoroughly discussed in the review article of Cowley \cite{Cowley80}, an incommensurate DW phase is characterized by a complex order parameter $\Delta_{\rm DW}=|\Delta|e^{i\phi}$ whose phase can take any continuous value. In the language of Landau theory of phase transitions, this corresponds to a phase-invariant free energy which depends only on the amplitude of the order parameter but not on its phase. The DW order at $T < T_{\rm DW}$ is then associated with a spontaneous symmetry breaking which fixes the phase of the order parameter and, according to Goldstone's theorem, gives rise to a gapless phason mode \cite{Bruce78c}. Physically, the phase of the DW order parameter characterizes the position of the electronic charge modulation with respect to the underlying lattice, and its arbitrariness, for an incommensurate DW state, corresponds to the translational invariance of the DW modulation. The soft phason mode can therefore be interpreted as a ``sliding mode'' \cite{Lee74} of the DW modulation with respect to the lattice, costing nearly no energy \cite{Lee74}. 

On the other hand, when the DW ordering vector is commensurate with the lattice spacing \cite{Lee74}, $Q=2\pi/pc_0$, with integer $p>2$ (in our case $p=4$ for the ($0,\uparrow$) level at $B_{\gamma}=48$ T), Umklapp scattering with the lattice potential generates $p$th order anharmonic terms in the free energy, of the form $|\Delta|^p \cos p\phi $, which explicitly break the phase invariance. The free energy thus depends now explicitly on $\phi$, so that the phase of the DW order parameter will be ``locked'' to one of the $p$ discrete values that minimize the free energy. In real space, the phase-locking mechanism corresponds to a ``pinning'' of the DW modulation peaks to the underlying lattice, with an energy gain due to the electron interaction with the lattice potential. In this case, continuous translational (or phase) invariance is lost, and the phason mode becomes gapped \cite{Sugai06}. 

\begin{figure*}[tb]
	\centering
		\includegraphics[width=1.0\textwidth]{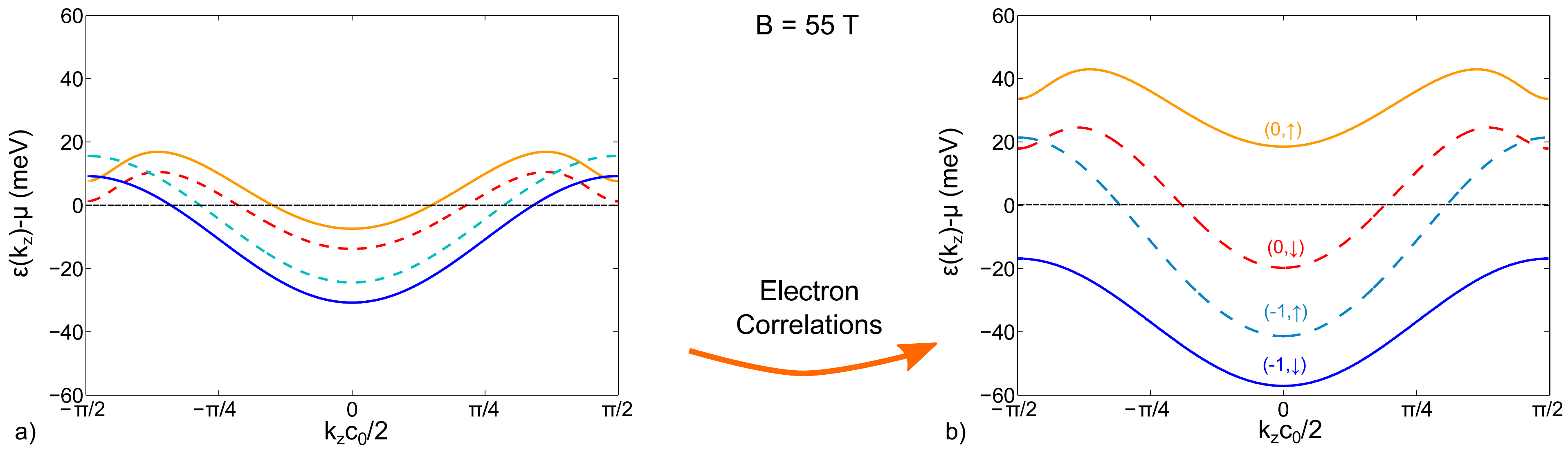}
	\caption{Effects of electron correlations on the LL-structure at high magnetic field ($B = 55$~T). a) Bare Slonczewski-Weiss-McClure tight-binding dispersion of the lowest LLs. b) Corresponding LLs with the inclusion of the electron self-energy correction, as calculated in our theoretical approach. Note the strong qualitative and quantitative effects of correlations, which lead to the emptying of the electron-like ($0,\uparrow$) and hole-like ($-1,\downarrow$) levels.}
	\label{fig:CorrelationsFigure}
%\vspace{1.2cm}
\end{figure*}

Using Landau theory, it can be shown quite generally that the ordering temperature of a commensurate DW is always lower than that of an incommensurate state. In fact, the phase-locking Umklapp energy dominates only at low temperatures, whereas at higher temperatures entropy favours the higher degeneracy of the incommensurate state. By lowering the temperature, therefore, one expects a lock-in transition, at $T_{\rm LIT} < T_{\rm CDW}$, from an incommensurate to a commensurate DW. This is precisely what we observe experimentally around 47.1 T, where $Q^{(0,\uparrow)} \approx 2\pi/4c_0$: An incommensurate DW order involving the $(0,\uparrow)$ Landau level sets in at $T_{\rm DW} \approx 10$~K, and subsequently undergoes a phase-locking transition into a commensurate order at $T_{\rm LIT} \approx 3$~K. The nature of this commensurate-incommensurate lock-in transition is rather subtle and requires a careful treatment. A naive application of Landau theory would in fact predict a first-order transition, as the free energies of the commensurate and incommensurate phases have different symmetry. However, this is only true if one considers a free energy with a spatially homogeneous order parameter. Instead, taking into account spatial variations of the order parameter (especially of its phase, which cost much less energy than amplitude fluctuations), it can be shown that the transition from the commensurate to the incommensurate state is described by the appearance and proliferation of domain walls separating regions of commensurate order \cite{McMillan76,Bak76,Bruce78b}. Across each of these domain walls, also known as ``solitons'' or discommensuration defects, the phase of the order parameter varies rapidly by $2\pi/p$, whereas it is nearly constant and locked to one of the commensurate values in the regions between solitons. In the vicinity of the lock-in transition the incommensurate phase is then characterized by a finite density of solitons and, as this density increases continuously from zero, the transition turns out to be of second order. Within this scenario, the gapless phason mode of the incommensurate phase appears as an acoustic mode corresponding to the oscillations of solitons about their mean positions \cite{Bruce78c}.

\subsection{5. Landau Level Calculation}

\begin{figure*}[tb]
\centering
\includegraphics[width=1.00\textwidth]{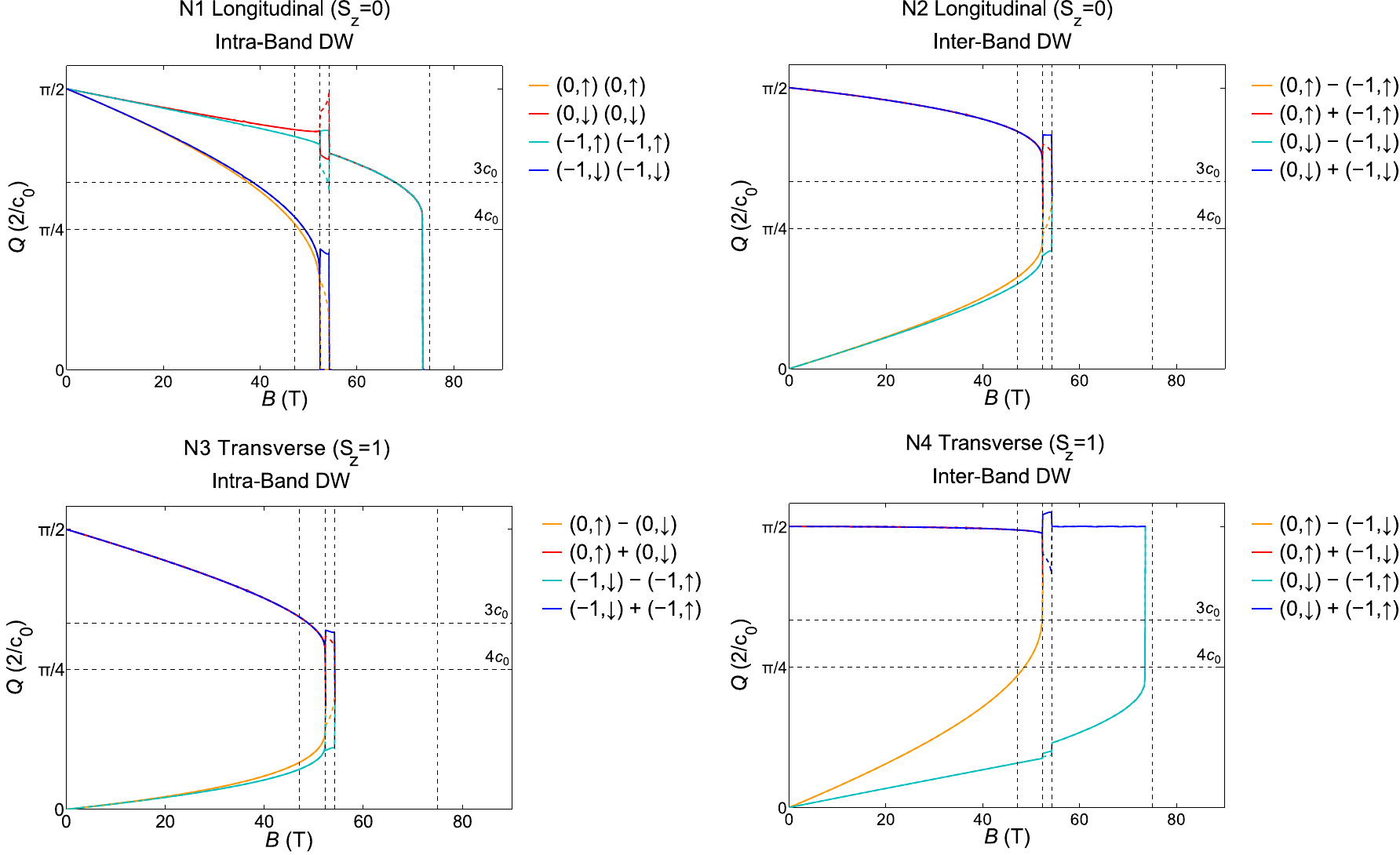}
\caption{Magnetic field dependence of the nesting vectors for the allowed intra- and inter-level nesting scenarios, as determined by our Landau level calculations. Solid lines correspond to nesting vectors in the H-K-H and dashed lines to nesting vectors in the H'-K'-H' valleys. In scenario N1 the nesting vectors are the usual $Q=2 k_F^{(n,\sigma)}$, whereas in scenarios N2 to N4 the nestings are given by $Q^{(\pm)}=k_F^{(n,\sigma)} \pm k_F^{(n',\sigma')}$. In the case of hole-like nesting vectors ($Q > \pi/c_0$) we have used $Q \to 2 \pi/c_0 - Q$. The vertical dashed lines correspond to the magnetic field values of the observed transitions, whereas the horizontal dashed lines indicate commensurate nestings. Note that the minus-sign combinations $Q^{(-)}=k_F^{(n,\sigma)} - k_F^{(n',\sigma')}$ are only included for the sake of completeness, but they do not contribute to the zero-temperature DW order parameters, due to Fermi blocking.}
\label{fig:NestingScenarios}
\end{figure*}

\noindent We calculate the renormalized energy dispersion 
$\epsilon_\nu({ k_z}) = \epsilon^{0}_\nu({ k_z}) + \Sigma_\nu({ k_z})$ 
of the relevant Landau levels (LLs) by including the 
electron self-energy correction $\Sigma_\nu({ k_z})$ to the bare 
Slonczewski-Weiss-McClure dispersion $\epsilon^{0}_\nu({ k_z})$ \cite{Slonczewski58,McClure60}. 
Here $\nu \equiv (n,\sigma,d)$ denotes the four doubly degenerate LLs,
with $n=-1,0$ (conventionally $n$ starts from $-1$ in graphite), $\sigma=\uparrow,\downarrow$, 
and the valley index $d$ accounting for the twofold 
degeneracy associated with the two inequivalent edges (H-K-H and $\rm H'$-$\rm K'$-$\rm H'$) of the Brillouin zone.
The momentum $k_z \in [-\pi/c_0, \pi/c_0]$ is measured along the Brillouin zone edges, where $c_0/2 = 3.354$ \AA\ is  
the distance between two neighboring graphene layers.
The electron self-energy is obtained in 
self-consistent Hartree-Fock approximation as
\begin{align}
\Sigma_\nu({ k_z}) & = \sum_{\nu', \, k_z', \, {\bf q}_\perp}
n_F^{\nu'}(k_z') \left[ {|F_{\nu \nu'}(0)|}^2 V_{\rm eff}(0) \right. \nonumber \\
 & \left. - \delta_{\nu \nu'} {|F_{\nu \nu'}({\bf q}_\perp)|}^2 V_{\rm eff}({\bf q}_\perp,k_z'-k_z) \right], \label{eq:self-energy}
\end{align}
where $n_F^{\nu}(k_z) = \Theta[\mu - \epsilon_\nu({ k_z})]$ is the zero-temperature Fermi distribution  
for the renormalized LLs, ${\bf q}_\perp$ is the momentum perpendicular to the $c$-axis 
(and hence to the magnetic field) measured from the Brillouin zone edge, and 
$F_{\nu \nu'}({\bf q}_\perp)$ are the appropriate form factors associated with the LLs: In the present 
calculation, $F_{\nu \nu'}({\bf q}_\perp) = \exp(-{\bf q}_\perp^2 l_B^2/4 )$ for $\nu$ and $\nu'$ 
belonging to the lowest LLs, where $l_B = \sqrt{\hbar c/ e B} $ is the magnetic length. 
The effective electron-electron interaction $V_{\rm eff}({\bf q}_\perp,q_z) \equiv V_{\rm eff}({\bf q})$
is calculated taking into account the effect of short-range correlations via the
so-called local-field correction (a concept first introduced by Hubbard\cite{Hubbard57}) 
to the random-phase approximation (RPA),
\begin{align}
V_{\rm eff}({\bf q}) & = V_0({\bf q}) - 
\frac{ \big[ (1-G_{{\bf q}}) V_0({\bf q}) \big]^2  \Pi({\bf q}; i0) }{ 1 + \big[ (1-G_{{\bf q}}) V_0({\bf q}) \big] 
 \Pi({\bf q}; i0) }. 
\end{align}
Here $V_0({\bf q})$ is the bare Coulomb potential,
$G_{{\bf q}}$ is the local-field correction factor 
(the standard RPA corresponds to $G_{{\bf q}} \equiv 0$), and
$\Pi({\bf q}; i\omega)$ is the one-loop electron polarization (particle-hole response function),
\begin{align}
\Pi({\bf q}; i\omega) & = - \frac{1}{2\pi l_B^2} \sum_{\nu} {|F_{\nu \nu}({\bf q}_\perp)|}^2 \nonumber \\
 & \times \int \frac{dk_z}{2\pi} \frac{ n_F^{\nu}(k_z) - n_F^{\nu}(k_z + q_z) }%
{i\omega + \epsilon_\nu({ k_z}) - \epsilon_\nu({ k_z + q_z})}.
\end{align}
Note that the polarization function
depends self-consistently on the renormalized LL dispersion. 
To obtain the local-field factor we use the approach developed by
Singwi, Tosi, Land, and Sj\"olander \cite{Singwi68} (STLS), which amounts
to solving the following self-consistent equations,
\begin{align}
G_{{\bf q}} & = \frac{1}{N_{\rm el}} \sum_{{\bf q}'} \frac{{\bf q} \cdot {\bf q}'}{{{\bf q}'}^2} 
 \left[ S({\bf q}+{\bf q}') -1 \right], \\
S({\bf q}) & = \frac{1}{N_{\rm el}} \int_0^\infty \frac{d\omega}{\pi} 
\frac{ \Pi({\bf q}; i\omega) }{ 1 + \big[ (1-G_{{\bf q}}) V_0({\bf q}) \big] 
 \Pi({\bf q}; i\omega) }, 
\end{align}
where ${N_{\rm el}}$ is the electron density and $S({\bf q})$ is the static structure factor.
We write the bare Coulomb potential as $V_0({\bf q}) = 4\pi e^2/(\varepsilon_r \varepsilon_0 {\bf q}^2)$,
where the relative permittivity $\varepsilon_r$ is taken as an ${\cal O}(1)$ adjustable parameter
which allows for the non-exact treatment of correlation effects.
The parameter $\varepsilon_r$ is fixed by adjusting the calculated value of $B_{\alpha'}$ to the corresponding experimental value, yielding $\varepsilon_r \approx 0.37$.
The renormalized chemical potential as a function of $B$ is determined by assuming the charge carrier balance (equal electron and hole densities) within the four LLs intersecting the Fermi energy.

In order to illustrate the prominent role of electron correlations in modifying the LL-structure, in Fig.~\ref{fig:CorrelationsFigure} we have plotted on the same scale the bare and renormalized LLs at a field of 55 T. Besides the quantitative effect of broadening the bandwidth of the innermost ($0,\downarrow$) and ($-1,\uparrow$) levels, electron correlations lead to the emptying of the two outermost LLs, thus accounting for the two first-order transitions $\alpha'$ and $\beta'$ observed between 52 and 55 T.

\subsection{6. Nesting Scenarios}

\noindent Here we would like to discuss the implications of our Landau level calculations on the possible high-field nesting instabilities.

In the magnetic quantum limit the LL-structure of graphite is characterized by two electron- and two hole-like LLs with opposite spin. Considering all inter- and intra-band density waves a total of ten nesting instabilities are possible. In general, each of these density-wave states will couple a given LL $(n,\sigma)$ with another level $(n',\sigma')$, yielding a translational symmetry breaking expectation value $\langle c^\dagger_{n\sigma,k_z+Q} c_{n'\sigma',k_z}^{\phantom{\dagger}} \rangle \neq 0$ ($c_{n\sigma,k_z}$ is the electron annihilation operator of the $(n,\sigma)$ LL). For each of these states, the corresponding nesting vectors are given by $Q^{(\pm)}=k_F^{(n,\sigma)} \pm k_F^{(n',\sigma')}$. Note that the $Q^{(-)}$ nesting vectors are considered here for the sake of completeness; However, the particle-hole expectation values associated with these vectors are actually vanishing at zero temperature due to the Fermi blocking mechanism, as discussed in more detail in the next section.
Assuming that all Landau level are either gapped by an inter- or intra-level nesting instability with a given spin symmetry, and that mixed scenarios do not occur, the number of nesting scenarios reduces significantly. 
Figure \ref{fig:NestingScenarios} shows the magnetic field dependence of the four remaining nesting scenarios which have a definite symmetry with respect to band-parity and spin: 
N1 describes longitudinal intra-level CDWs (even-parity, even-spin); N2 longitudinal inter-level CDWs (odd-parity, even-spin); N3 transverse intra-level spin-density waves (even-parity, odd-spin); N4 transverse inter-level spin-density waves (odd-parity, odd-spin). 

Combining all the experimental observations from our and other groups \cite{Fauque13,Zhu15} recent data, the allowed nesting scenarios are then only those that are compatible with the following observed features: i) low-temperature lock-in transition $\gamma$ around 47 T, due to commensurate nesting; ii) sharp first-order transitions $\alpha'$ and $\beta'$ at 52.3 and 54.2~T; iii) sharp first-order transition $\delta'$ around 75 T. If we assume the dominance of a single nesting channel across the entire phase diagram, so that all density waves are of the same kind, then the only possibility of combining all the experimental features is given by the N1 scenario, i.e., the longitudinal intra-band CDW order proposed in the main text. 
Alternative scenarios are possible if we lift this assumption and consider the possibility of different dominant instabilities  in different regions of the phase diagram. In this case, in the region below 55 T (more precisely, below the $\beta'$ transition) we could have two transverse intra-band spin-density waves (N3), with nesting vectors $Q^{(+)}$ commensurate with $3c_0$ around 47~T, whereas above 55 T an alternative nesting scenario to N1 could be the N4, with a single inter-band transverse spin-density wave accounting for the high-field density wave state observed by Fauque \textit{et al.} \cite{Fauque13} and Zhu \textit{et al.} \cite{Zhu15}. 

\begin{figure}[tb]
	\centering
		\includegraphics[width=1.0\columnwidth]{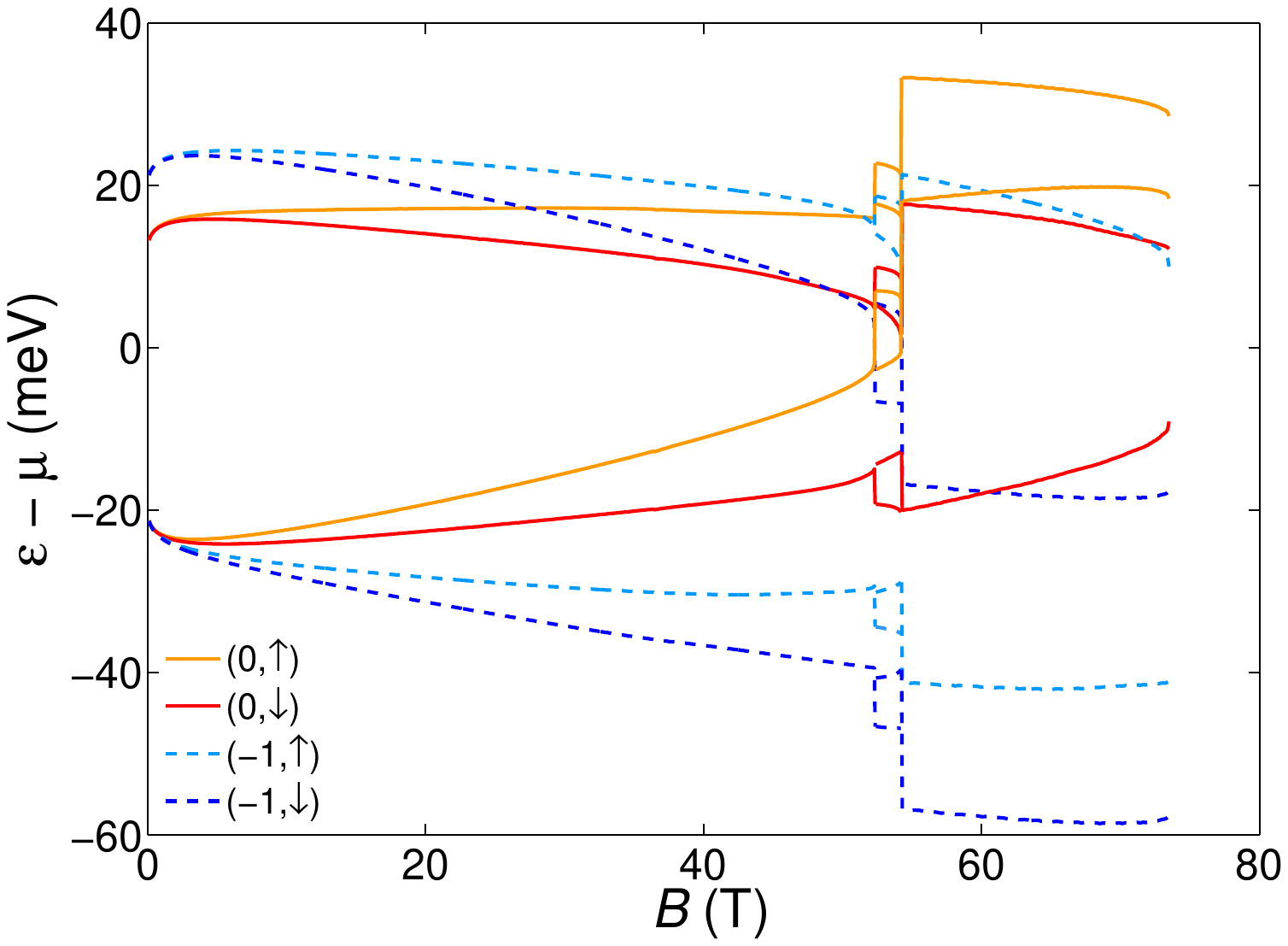}
	\caption{Band maxima and minima of the Landau levels: Solid (dashed) curves correspond to 
	electron-like (hole-like) LLs.}
	\label{fig:MaxANDMins}
\end{figure}

In addition, we would like to comment on the possibility that the observed density-wave states are driven by an excitonic insulator instability rather than a Fermi-level nesting instability. Excitonic insulators are formed when an electron and a hole from nearly overlapping bands form a bound pair. In order for this mechanism to take place, the overlap between the bands should be sufficiently small and of the order of magnitude of the resulting gap opening, which from the out-of-plane resistance anomalies turns out to be of the order of 1~$\rm meV$. Based on our calculations, this gap energy scale is much smaller than the energy differences between electron bands minima and hole bands maxima, except possibly in a narrow region below the $\beta'$ transition (see Fig. \ref{fig:MaxANDMins}). Since the resistance anomalies take place over a much wider range of magnetic fields, it is rather unlikely that the excitonic insulator mechanism is the driving force responsible for the electronic phases in both the low-field (below 55 T) and the high-field region (between 55 and 75 T) of the phase diagram.

\subsection{7. Density wave order parameters}

\noindent Following the previous discussion on the possible nesting scenarios, we have calculated the corresponding density wave (DW) order parameters associated with each of the four scenarios, N1 to N4. The information on the amplitude of the density wave gaps helps us to resolve the ambiguities among the multiple scenarios that are compatible with the observed experimental facts, e.g., the N1 vs N3 dichotomy in the region below the $\beta'$ transition, and the N1 vs N4 in the region above it.

We decouple at Hartree-Fock level the effective density-density Coulomb interaction,
\begin{align}
H_{\rm int} & = \frac12 \sum_{{\bf q}} V_{\rm eff}({\bf q}) \rho_{\bf q} \rho_{-\bf q},
\end{align} 
considering all possible DW-channels with nesting vectors within the lowest four (doubly degenerate) LLs $\nu \equiv (n,\sigma)$. This yields the following Hartree-Fock interacting Hamiltonian for the one-dimensional LLs,
\begin{align}
H_{\rm int}^{\rm HF} & = \sum_{\nu, \, \nu', \, k_z} \Delta_{\nu' \nu}(k_z) 
c^\dagger_{\nu',k_z - Q_{\nu \nu'}} c_{\nu,k_z}^{\phantom{\dagger}} + {\rm H. c.},
\end{align} 
where $Q_{\nu \nu'} = k_F^{(\nu)} + k_F^{(\nu')}$ are the DW nesting vectors and
\begin{align}
\Delta_{\nu' \nu}(k_z) & = - \sum_{{\bf q}_\perp, \, k_z'} {|F_{\nu \nu'}({\bf q}_\perp)|}^2
 V_{\rm eff}({\bf q}_\perp, k_z - k_z') \nonumber \\
 & \times \langle c^\dagger_{\nu,k_z'} c_{\nu',k_z' - Q_{\nu \nu'}}^{\phantom{\dagger}} \rangle 
\label{eq:gap-eq}
\end{align} 
are the corresponding self-consistent DW order parameters. In defining the nesting vectors we have not considered the minus sign combinations, $Q_{\nu \nu'}^{(-)} = k_F^{(\nu)} - k_F^{(\nu')}$, because these combinations do not give rise to finite order parameters. In other words, $\langle c^\dagger_{\nu,k_z} c_{\nu',k_z - Q_{\nu \nu'}^{(-)}} \rangle \equiv 0$. The reason for this is the Fermi blocking mechanism, which at zero temperature yields a vanishing expectation value of a particle-hole pair whose states are either both empty or both occupied. To see how the mechanism works, assume, for example, $k_F^{(\nu)} > k_F^{(\nu')} > 0$. Then for $k_z$ in the vicinity of $k_F^{(\nu)}$, say $k_z = k_F^{(\nu)} + dk$ (with $|dk| \ll k_F^{(\nu)}$), and $Q_{\nu \nu'}^{(-)} = k_F^{(\nu)} - k_F^{(\nu')}$, we have $k_z - Q_{\nu \nu'}^{(-)} =  k_F^{(\nu')} + dk$. That means that, regardless of the sign of $dk$, both states $|\nu,k_z\rangle$ and $|\nu',k_z - Q_{\nu \nu'}^{(-)}\rangle$ are either empty or occupied: If $dk$ is positive, they are both empty, if it is negative, they are both occupied. Obviously this works also around the other $k_F$ points. We have explicitly checked that this mechanism is at work in our self-consistent calculation.

The full Hartree-Fock Hamiltonian can now be written in the following matrix form,
\begin{align}
H^{\rm HF} & = \sum_{\nu, \, \nu', \, k_z, \, k_z'} c^\dagger_{\nu',k_z'} 
\hat{D}_{\nu'k_z', \, \nu k_z}^{\phantom{\dagger}} c_{\nu,k_z}^{\phantom{\dagger}}, \label{eq:Ham_D} \\
\hat{D}_{\nu'k_z', \, \nu k_z}^{\phantom{\dagger}} & = \delta_{\nu',\nu} \delta_{k_z',k_z} 
\left[ \epsilon_\nu({ k_z}) - \mu \right] \nonumber \\
 & + \left[ \Delta_{\nu' \nu}(k_z) \delta_{k_z',k_z - Q_{\nu \nu'}} + {\rm H. c.} \right].
\end{align} 
The complete energy spectrum is given by the eigenvalues $E_i$ of the $\hat{D}$-matrix,
$(\hat{U}^\dagger \hat{D} \hat{U})_{ij} = E_i \delta_{ij}$, with corresponding eigenvectors 
$\hat{U}_{\nu k_z, i}$. By numerically diagonalizing the $\hat{D}$-matrix one finally obtains the expectation values 
\begin{align} 
\langle c^\dagger_{\nu,k_z} c_{\nu',k_z'}^{\phantom{\dagger}} \rangle & =
\sum_i \hat{U}_{\nu' k_z', i} \hat{U}^*_{\nu k_z, i} n_F(E_i) 
\end{align} 
that enter the self-consistent gap equation (\ref{eq:gap-eq}). Here $n_F(E_i)$ denotes the Fermi function, which at zero-temperature is given by $n_F(E_i)=\Theta(-E_i)$. The Fermi function used here differs from the one used in Eq.~(\ref{eq:self-energy}) because the chemical potential is already included in the diagonal part of the Hamiltonian (\ref{eq:Ham_D}). Note that, for a given channel, the definite symmetry simplifies greatly the general form of the $\hat{D}$-matrix, which acquires a block-diagonal structure. Each block $\hat{D}^{\ell}$ can be diagonalized independently and corresponds to a given DW band with energy spectrum $E_i^{\ell}$. A finite DW order parameter gives rise to a spectrum characterized by two continuous branches, with positive and negative energies, separated by an energy gap. We can therefore define the spectral gap associated with the DW $\ell$-band as
\begin{align} 
\Gamma_{\ell} & = \frac12 
\left[ \min_{i: E_i^{\ell}>0} (E_i^{\ell}) - \max_{i: E_i^{\ell}<0} (E_i^{\ell}) \right].
\end{align}
Together with the definition of the spectral gaps, we also introduce the thermodynamic order parameters $\Delta_{\ell}$, defined as the average over the Brillouin zone of the $k_z$-dependent DW order parameters:
\begin{align}
\Delta_{\ell} & = \int_{-\pi/c_0}^{\pi/c_0} \frac{dk_z}{(2\pi/c_0)} \Delta_{\nu' \nu}(k_z).
\end{align}
Here we have used $\ell \equiv (\nu',\nu)$ in order to identify a given DW channel with nesting between levels $\nu$ and $\nu'$.
While the spectral gaps are the quantities that are directly related to the modifications in the transport properties, the thermodynamic order parameter gives a more direct information on the strength of a given DW instability and therefore is more closely related to the DW critical temperature.

These calculated density wave gaps must be treated with caution, since they depend exponentially on the coupling strength, the spin dependence of which has not been considered so far in this work (and will be the subject of future work). This is analogous to the well-known difficulty of reliable \textit{a priori} calculations of superconducting transition temperatures/gaps in BCS theory.
In particular, in the low-field limit it is crucial to include the full hierarchy of LLs crossing the Fermi energy in order to recover the exponential suppression of the order parameter in the limit $B \to 0$. 
Nevertheless, these results 
%probably 
give a reasonable indication of the hierarchy of gaps in a given scenario (e.g. N1, CDW) and for fields larger than 20 T, where the approximation of taking into account only the lowest LLs is well justified. They also show how the gap responds to the commensuration transition, or proximity to it.

\subsection{8. Density Wave Scenarios}

\noindent We now turn to the discussion on which of the four scenarios with definite symmetry with respect to band-parity and spin describes best our experimental findings, using the thermodynamic order parameter as the reference quantity to describe the tendency toward a given DW instability.

\begin{figure*}[tb]
\begin{center}
	\includegraphics[width=0.9\textwidth]{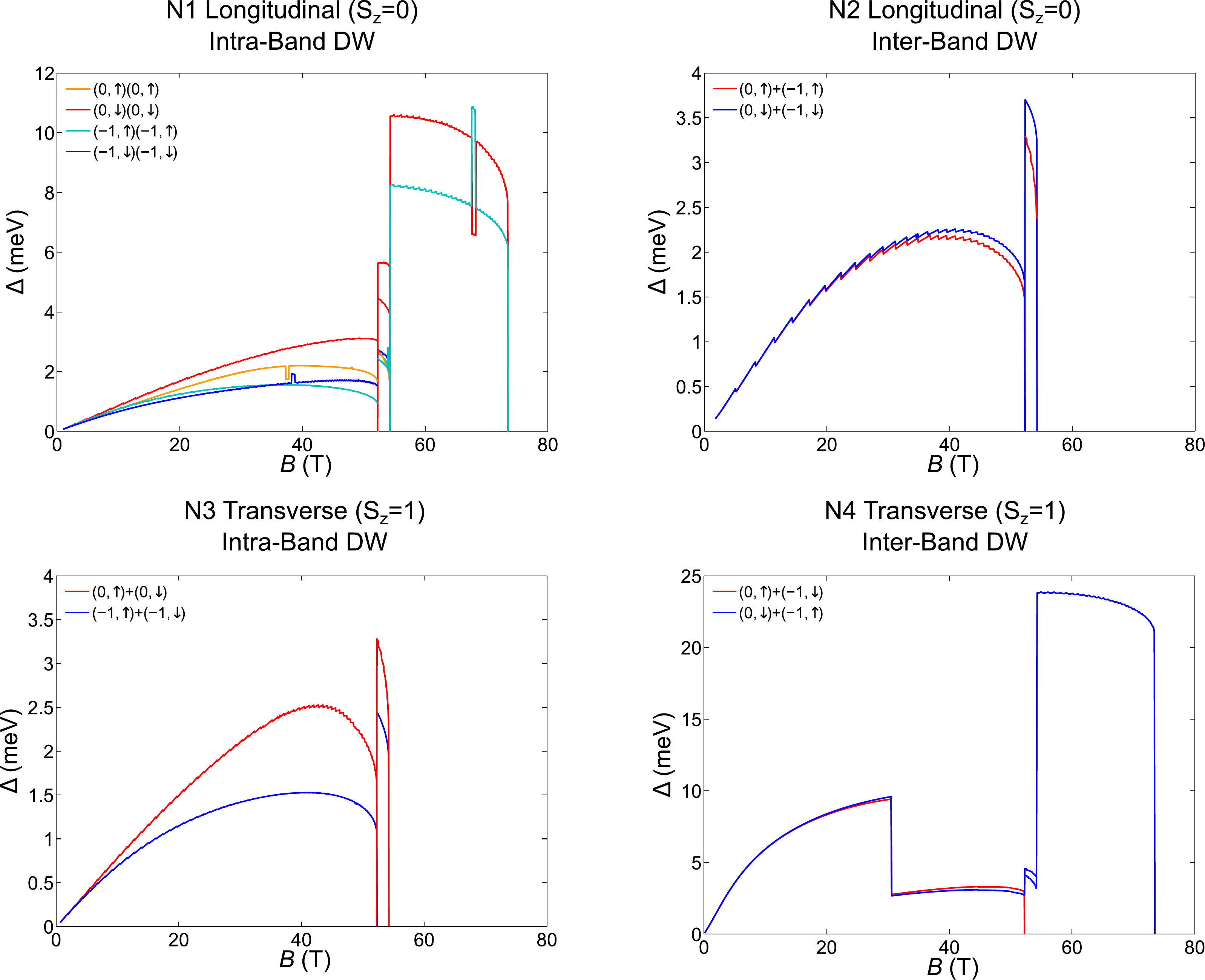}
\end{center}
\caption{DW order parameters for different nesting scenarios within the four lowest Landau levels: (a) intra-band even-spin DWs, (b) inter-band even-spin DWs, (c) intra-band odd-spin DWs, (d) inter-band odd-spin DWs.}
\label{fig:Gaps}
\end{figure*}

\begin{enumerate}
\item
In scenario N1 (intra-band, even-spin) we can have up to four DW bands, with ${\ell} = (0\uparrow, 0\uparrow)$, $(0\downarrow, 0\downarrow)$, $(-1\uparrow, -1\uparrow)$, $(-1\downarrow, -1\downarrow)$. 
We find pronounced commensuration effects with $3 c_0$ appearing in the field dependence of the order parameter at the field values where the Q-vectors are equal to $2\pi/(3 c_0)$ (see Fig \ref{fig:NestingScenarios}). These are seen at 37.3-37.8 T in the order parameter of the $(0 \uparrow, 0 \uparrow)$ DW band, at 38.2-38.8 T in the $(-1\downarrow, -1\downarrow)$ DW band, 
and around 67.6-68.1 T in the $(0 \downarrow, 0\downarrow)$ and $(-1\uparrow, -1\uparrow)$ DW bands, whose nesting vectors are equal in the region above 54.2 T due to charge neutrality (see Fig. \ref{fig:NestingScenarios}).
A weaker commensuration effect with $4 c_0$ is observed at 47.9-48.1 T in the $(0\uparrow, 0\uparrow)$ DW band. Perhaps an even weaker $4 c_0$ commensuration effect could take place in the $(-1\downarrow, -1\downarrow)$ DW band at 49.2 T, but it is beyond the present numerical accuracy of our calculations. 
In addition, in the $\alpha'$-$\beta'$ split region between 52.3 and 54.2 T, where we predict a valley-density-wave order, we observe another $3 c_0$ commensuration effect in the $(-1\uparrow, -1\uparrow)$ DW band of the electron-doped valley.
The largest order parameter is found to be the $(0\downarrow, 0\downarrow)$ over the entire magnetic field range, whereas the other DW instabilities are close to each other in the region where they can coexist, i.e., below 54.2 T.

\item
In scenario N2 (inter-band, even-spin) we can have two DW bands, with ${\ell} = (0 \! \uparrow,\! -1 \! \uparrow)$ and $(0 \! \downarrow,\! -1 \! \downarrow$). Here the Q-vectors do not take commensurate values, and hence the field dependence of the gaps turns out to be smooth.
The gaps are very close to each other, with a slightly larger value of the $(0\downarrow, -1\downarrow)$ one.

\item
In scenario N3 (intra-band, odd-spin) we can have two DW bands, with ${\ell} = (0 \! \uparrow, 0 \! \downarrow)$ and $(-1 \! \uparrow, -1 \! \downarrow)$. Here, interestingly enough, we do not observe commensuration effects in the order parameters, although at 48.7 T the Q-vectors (equal to each other due to charge neutrality) are equal to $2\pi/(3 c_0)$. The understanding of this absence should be addressed in a forthcoming publication, focused on the temperature dependence of the DW order parameters.
We find the $(0 \uparrow, 0\downarrow)$ order parameter almost as twice as large as the $(-1\uparrow, -1\downarrow)$ one.

\item
Scenario N4 (inter-band, odd-spin) is somehow anomalous because its nesting vectors are almost field-independent, and almost equal to the commensurate value $\pi/c_0$. More precisely, in the region below 52.3 T the Q-vectors decrease extremely slowly from the zero-field value of $\pi/c_0$, whereas they are exactly equal to $\pi/c_0$ in the region between 54.2 and 73.5 due to charge neutrality. 
The two order parameters, ${\ell} = (0\uparrow, -1\downarrow)$ and $(0\downarrow, -1\uparrow)$, are almost identical and are found to be very large whenever Q is sufficiently close to the commensurate value, namely below 30 T and in the whole region between 54.2 and 73.5. At approximately 30 T, a transition from a $2 c_0$ commensurate DW to an incommensurate DW should take place. It is worth noting that a $2 c_0$ odd-spin commensurate DW corresponds, in fact, to an antiferromagnetic order.

\end{enumerate}

\subsection{9. Comparison with Experiment}

\noindent Comparing these theoretical findings with our experimental observations imposes strict constraints on the possible DW instabilities, which we now discuss.

We believe that N1 offers the most likely DW scenario throughout the magnetic field range up to $75\,\mathrm{T}$, above which no DW order can take place in the bulk due to absence of any Fermi-level nesting. Let us discuss in more detail the reasoning that led us to this conclusion. 

Consider first the region below $52.3\,\mathrm{T}$ ($\alpha'$ transition).
Scenario N2 offers no commensurate values for the Q-vectors that could result in a DW lock-in transition. In particular, the order parameter does not show any feature around $47.1\,\mathrm{T}$, where experimentally we observe a sharp resistance anomaly below $3\,\mathrm{K}$, interpreted as a lock-in transition of the DW order. 

Scenario N3 could in principle represent a valid alternative to N1 below $52.3\,\mathrm{T}$, considering that the Q-vectors are commensurate with $3 c_0$ at $48.7\,\mathrm{T}$, i.e., not too far from the experimental value where the lock-in transition is observed. However, the order parameters of the two DW bands do not show any sizeable commensuration effect that can hint for a lock-in transition at this field value: If commensuration effects are present, they are probably below the numerical resolution of our calculations, and in any case smaller than the one characterizing the N1 $(0 \uparrow, 0 \uparrow)$ DW band at $48\,\mathrm{T}$.  

In scenario N4 the theory predicts a very strong transition around 30 T, where the DW order switches from an antiferromagnetic $2 c_0$ commensurate state to an incommensurate spin-density wave. We are led to exclude this scenario for the following reasons: (i) the strong transition at $30\,\mathrm{T}$, predicted by the theory, is not observed experimentally; (ii) the antiferromagnetic order, predicted below $30\,\mathrm{T}$, is likely to produce a strong signal in several response functions and, if present, it should have been detected by various experimental techniques, considering that it develops in the well-studied low field region; (iii) no commensurate effects are present around $47.1\,\mathrm{T}$ that could explain the resistance anomaly observed below $3\,\mathrm{K}$; (iv) there is no mechanism to explain the $\beta'$ feature because the $(0 \uparrow, -1 \downarrow)$ DW collapses already at $\alpha'$, with the emptying of the $(0  \uparrow)$ and $(-1 \downarrow)$ levels in different sub-valleys, whereas the $(0 \downarrow, -1 \uparrow)$ DW survives across the $\beta'$ transition.

We note that the calculated gaps in the N4 scenario below $30\,\mathrm{T}$ exceed the CDW gaps (N1 scenario) by almost an order of magnitude. Since we have already ruled out the N4 scenario below $54\,\mathrm{T}$, this suggests that the spin dependence of interactions disfavors SDW scenarios (odd spin, N3 and N4), and requires more theoretical work. However this implies that the non-observation of the $(0\uparrow$,~$-1\downarrow)$ SDW (N4) would also rule out both $(0\uparrow, 0\downarrow)$ and $(-1\uparrow, -1\downarrow)$ SDWs (N3) below $54\,\mathrm{T}$, which in our approximation have comparable gaps to the intraband CDWs, $(0\uparrow, 0\uparrow)$ and $(-1\downarrow, -1\downarrow)$ (N1).

In summary, we experimentally resolve the collapse of two DWs, at $52.3\,\mathrm{T}$ and $54.2\,\mathrm{T}$. Theoretically, the $(0\uparrow)$ and $(-1\downarrow)$ levels empty at these fields. In the light of the discussion above we propose that below $52\,\mathrm{T}$ there exist at least two DW states, corresponding to the $(0\uparrow$,~$0\uparrow)$ and $(-1\downarrow, -1\downarrow)$ CDWs (N1), to account for the observations so far.

We predict a $3c_0$ lock-in transition in both the $(0\uparrow, 0\uparrow)$ and $(-1\downarrow, -1\downarrow)$ CDWs at close by fields around $38\,\mathrm{T}$, which we did not observe.
The lack of this feature in the current data might be due to insufficient resolution at low temperatures (below 2 K) and the concomitant presence of the entrant $\beta$-transition at the same temperature where the lock-in transition in the $(0 \uparrow, 0 \uparrow)$ DW would take place. In fact, note that the lock-in transition temperature is itself significantly lower than the onset temperature of the DW order: In the case of the $\gamma$ transition at $47.1\,\mathrm{T}$, for example, we have $T_{\rm DW} \approx 10\,\mathrm{K}$ and $T_{\rm LIT} \approx 3\,\mathrm{K}$. Accordingly, we expect a lock-in transition in the $(0 \uparrow, 0 \uparrow)$ DW at $38\,\mathrm{T}$ to appear only at temperatures around $1.5\,\mathrm{K}$, where the $\beta$-transition signature takes place. 
Theoretically we also find that the $3 c_0$ commensuration has an opposite effect on the DW gaps. Therefore it could give rise to compensating signatures in resistance, making them hard to observe. This issue would potentially be resolved by a high-resolution field study of these lock-in transitions by a technique designed to be sensitive to the concomitant lattice distortion that would be expected.

The results of Fauque \textit{et al.} demonstrate that DWs exist above $54\,\mathrm{T}$ up to $75\,\mathrm{T}$ where they collapse. As discussed in the main manuscript our calculations confirm the field ($75\,\mathrm{T}$) at which the collapse occurs through emptying both the $(0\downarrow)$ and $(-1\uparrow)$ LLs. The $(0\downarrow,-1\uparrow)$ interband SDW (scenario N4) is a candidate in this field regime. The non-observation of the $(0\uparrow,-1\downarrow)$ SDW (N4) below 54 T, as discussed, points to disfavoring SDWs. However the $(0\downarrow,-1\uparrow)$ SDW must still remain a possible candidate between $54\,\mathrm{T}$ and $75\,\mathrm{T}$.

The only remaining possibilities in the field range $54\,\mathrm{T}$ to $75\,\mathrm{T}$ are intraband CDWs $(0\downarrow, 0\downarrow)$ and $(-1\uparrow, -1\uparrow)$. We find that these have comparable gaps. The gaps of these two DWs jump up significantly at $54\,\mathrm{T}$ with the emptying of the $(0\uparrow)$, $(1\downarrow)$ LLs. We therefore propose a coexistence of these two CDWs over this field range.

Importantly we predict a sharp commensuration feature in the $(0\downarrow, 0\downarrow)$ and $(-1\uparrow, -1\uparrow)$ DWs at $68\,\mathrm{T}$ (see Fig.~\ref{fig:Gaps}). There may be hints of this in the resistance data of Fauque. Confirmation of this commensuration by another technique would confirm our proposal. In principle, and more challengingly, further support would also be provided by resolution of the onset of $(0\downarrow, 0\downarrow)$ and $(-1\uparrow, -1\uparrow)$ DWs sequentially in this field regime.

We expect the hierarchy of gaps within a given scenario (eg N1, CDW) to be reliable. These calculations show that the gaps of the $(0\downarrow, 0\downarrow)$ and $(0\uparrow, 0\uparrow)$ CDWs are close below 30T, and somewhat larger than those for the $(-1\downarrow, -1\downarrow)$ and $(-1\uparrow, -1\uparrow)$ CDWs. This suggests that all four CDWs are present below $52\,\mathrm{T}$, given that the onset temperatures of CDWs in spin split LLs would be hard to resolve. The gaps also support the hierarchy in which the CDW in the electron LL occurs first with decreasing temperature.

\begin{figure}[tb]
	\centering
		\includegraphics[width=0.95\columnwidth]{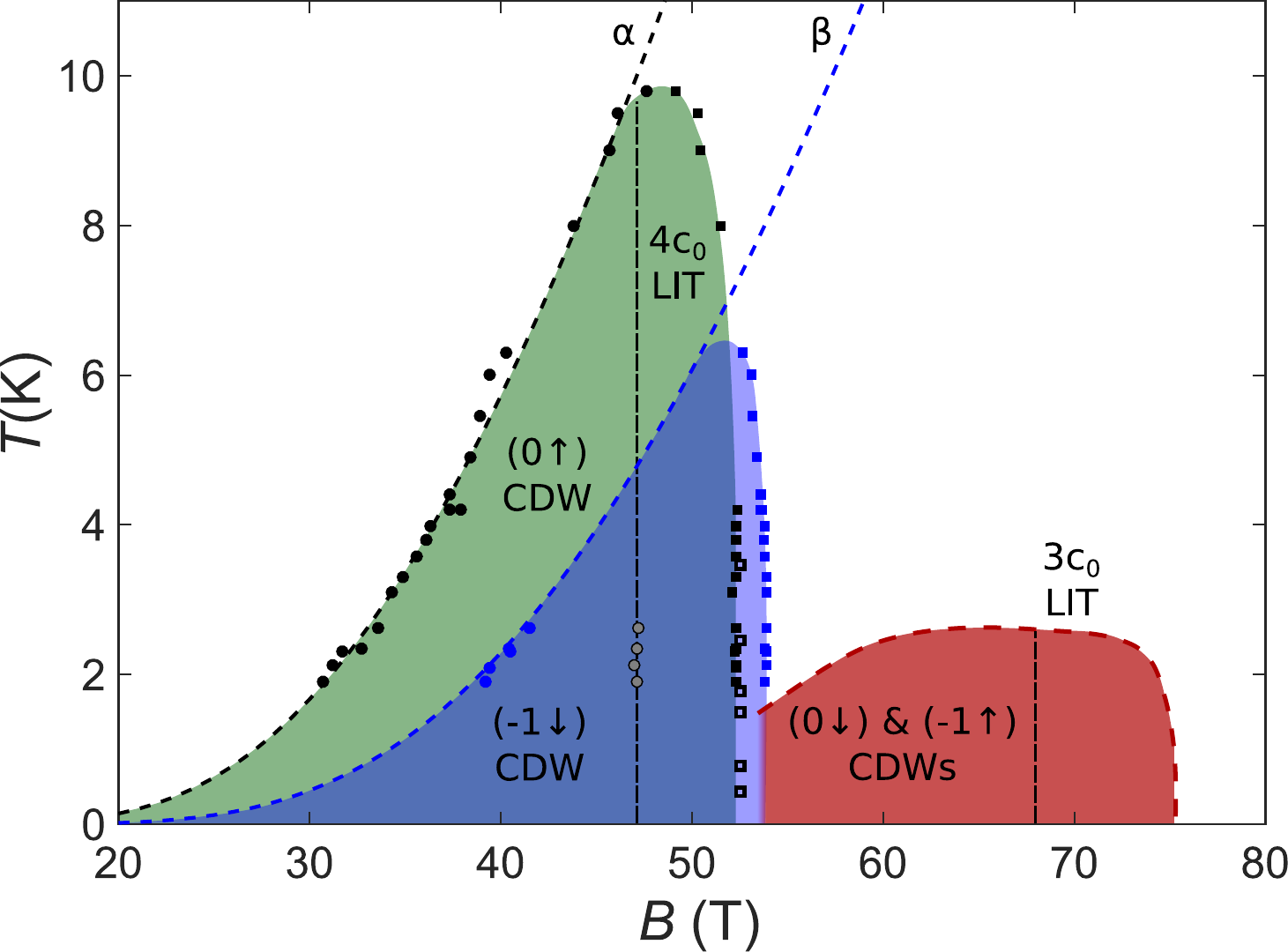}
	\caption{Proposed schematic phase diagram of the CDWs in graphite. Circles and squares are phase transitions with positive and negative differential magnetoresistance as determined from in-plane resistances, where black represents the $\alpha$, blue the $\beta$-transition. At low fields the critical temperature of $\alpha$ and $\beta$ increases exponentially  following Yoshioka-Fukuyama's theory \cite{Yoshioka81} (black and blue dashed lines): $T_c(B) = T\times \exp(-B*/B)$, with $T_\alpha = 230\,\mathrm{K}$, $B_\alpha = 148\,\mathrm{T}$, $T_\beta= 300\,\mathrm{K}$ and $B_\beta = 195\,\mathrm{T}$. The red colored high field phase has been extracted from the out-of-plane resistance shown in Fauque \textit{et al.} \cite{Fauque13}.}
	\label{fig:PhaseDiagram}
\end{figure}

This logic leads to the most likely, and also appealing, scenario, that all four LLs are subject to CDW instabilities (see above for caveats). Two of these survive above $54\,\mathrm{T}$ before collapsing at $75\,\mathrm{T}$. How the onset of these $(0,\downarrow)$ and $(-1,\uparrow)$ instabilities track through the emptying of the LLs at $54\,\mathrm{T}$ is an open question. A proposed phase diagram is shown in Fig. \ref{fig:PhaseDiagram}.  This shows the observed and predicted commensuration transitions. Observation of the predicted lock-in transition at $68\,\mathrm{T}$ would confirm this hypothesis. Probes that detect lock-in transitions via sensitivity to lattice distortion would provide additional support for the CDW proposal.

It should be mentioned that the phase diagram Fig.~\ref{fig:PhaseDiagram} is based on magnetic field derivatives of the in-plane (below 54 T) and out-of-plane (above 54 T) resistance. Hence, caution must be taken when comparing the critical temperatures on the two sides of the phase diagram. Generally, activated behavior is only observed in $\rho_\mathrm{zz}$ when all LLs are gapped, i.e., there is no metallic conduction originating from ungapped LLs which would otherwise short-circuit the out-of-plane conductance. For this reason, the transition from metallic to activated behavior in $\rho_\mathrm{zz}$ is sensitive to the smallest gap in the system.  
Transitions between partially gapped states, such as $\alpha$, are nonetheless observable in the in-plane resistance. Due to orbital effects, the transverse conductivity in high mobility metals drops rapidly in magnetic field following $\sigma_\mathrm{xx}(B)\approx\sum_{e,h}\sigma^{e,h}_\mathrm{xx}(0)/(1+\omega^2_\mathrm{c}\tau^2)$ where $\omega_c$ and $\tau$ are the electron and hole cyclotron frequency and scattering time, whilst $\sigma_\mathrm{zz}(B)\approx \mathrm{const}$ \cite{Pippard}. It follows that at highest fields $\sigma_\mathrm{xx}\ll\sigma_\mathrm{zz}$, making $\sigma_\mathrm{xx}$ sensitive to small conductivity changes as originating from activated LLs. In other words, short-circuiting of gapped LLs in $\sigma_\mathrm{xx}$ is suppressed as compared to $\rho_\mathrm{zz}$. However, as pointed out in the main article, we also note that above 54 T highly conductive ballistic edge states appear, rendering the in-plane resistance insensitive to bulk gaps.

At $\alpha$ the system goes into a partially gapped state with CDWs in the $(0\uparrow)$ and $(0\downarrow)$ LLs, whereas $\beta$ marks the field at which all four LLs are gapped. Consequentially the observation of activated behavior in the out-of-plane resistance is direct evidence for DWs in all four LLs. In this regime electrical transport is governed by activation across the smallest gap in the system, i.e. the CDW gaps in the $(-1\uparrow)$ and $(-1\downarrow)$ LLs. The direct determination of these gaps, e.g. by Arrhenius plots, is complicated by impurity and surface conductivity leading to a resistance saturation at lowest temperatures.

Furthermore, we account for the absence of the lock-in transition $\gamma$ in the out-of-plane transport by its sensitivity to the lowest DW gap. As shown in Fig. \ref{fig:NestingScenarios}, commensuration takes place in the $(0\uparrow)$ and $(-1\downarrow)$ LLs. However, the gaps associated with these LLs are not the smallest in the system (see Fig. \ref{fig:Gaps} N1). Thus a change of the subdominant gaps does not cause a change of the activated out-of-plane resistance.

Throughout the high field CDW dome between $54$ and $75\,\mathrm{T}$ \cite{Fauque13} (Fig. \ref{fig:PhaseDiagram}) activated behavior is observed, indicating a phase where both the $(0\downarrow)$ and $(-1\uparrow)$ LLs are gapped by CDWs. However, from Fig. \ref{fig:Gaps}, we know that the CDW gaps in both of these LLs differ significantly. Thus it must be assumed the $(0\downarrow)$ CDW dome extends far beyond the phase boundary shown in Fig. \ref{fig:PhaseDiagram}.


\begin{thebibliography}{61}
\bibitem [{dop()}]{doping}%
  \BibitemOpen
  \href@noop {} {}\bibinfo {note} {A uniform charge imbalance of 0.11 is
  required to explain the observed splitting, which is unphysical
  \cite{Akiba15}.}\BibitemShut {Stop}%
\end{thebibliography}
\end{document}